\newcommand{\PaperTitle}{Quality of Coverage (QoC):   Quantifying Cellular Network Coverage Quality, Usability and Stability}
\documentclass[10pt,sigconf,letterpaper, nonacm]{acmart}

\usepackage{amsmath, amssymb, graphicx}
\usepackage{hyperref}

\AtBeginDocument{%
  }

\setcopyright{none}
\renewcommand\footnotetextcopyrightpermission[1]{} 
\usepackage[english]{babel}
\usepackage{graphicx}
\usepackage{graphics}
\usepackage{xcolor}
\usepackage[utf8]{inputenc}
\usepackage{amsmath}
\usepackage{tabularx}
\usepackage{array, multirow} 
\usepackage{subcaption}
\usepackage{caption}
\usepackage{enumitem}
\usepackage{wrapfig}
\usepackage{textcomp}

\newcommand{\emb}[1]{\textit{\textcolor{purple}{[EMB]: #1}}}

\newcommand{\ignore}[1]{}

\begin{document}

\title[Quality of Coverage (QoC)]{\PaperTitle}

\author{Varshika Srinivasavaradhan}
\email{varshika@ucsb.edu}
\affiliation{%
  \institution{University of California, Santa Barbara}
  \country{USA}
}

\author{Morgan Vigil-Hayes}
\email{vigilhay@msu.edu}
\affiliation{%
  \institution{Michigan State University}
  \country{USA}
}

\author{Ellen Zegura}
\email{ewz@cc.gatech.edu}
\affiliation{%
  \institution{Georgia Institute of Technology}
  \country{USA}
}

\author{Elizabeth Belding}
\email{ebelding@ucsb.edu}
\affiliation{%
  \institution{University of California, Santa Barbara}
  \country{USA}
}

\begin{abstract}

Characterizing cellular network performance is complex. Current representations of cellular coverage, such as service provider and FCC coverage maps, focus only on the minimal level of available bandwidth (e.g., 35/3Mbps download/upload speed for 5G) and omit critical dimensions of quality: network usability and stability over space and time.  Because cellular  performance can vary substantially along both dimensions, a more fine-grained characterization is necessary.  
We introduce Quality of Coverage (QoC), a novel multi-dimensional set of key performance indicators  (KPIs) that capture measured temporal and spatial  
performance quality, usability and stability.   To evaluate QoC, we first analyze whether the QoC KPIs accurately reflect expected network behavior at individual locations and across spatially-aggregated regions.  Then, we apply QoC to more than 15~million measurements from a production network to  evaluate its ability to characterize  real-world network behavior. Together, our results demonstrate the need for KPIs that capture the full spectrum of cellular performance and show how QoC enables rigorous evaluation of coverage quality across multiple geographic scales.
\end{abstract}

\maketitle

\section{Introduction}
Networking research and practice has long used path-level metrics 
to characterize and improve network quality (e.g.~\cite{paxson1996routing,Paxson1997:End}), to help users understand 
what they can expect from network performance, and to let providers express what they will deliver
to customers. These metrics collectively are referred to as Quality of Service (QoS) and include 
measures such as download and upload speed, end-to-end latency, jitter, and path loss. QoS metrics 
for  fixed network characterization are well understood and widely used. 

Quality of Experience (QoE) metrics are relatively more recent and meant to fill a gap left by QoS, 
namely to better connect network performance to user-experienced application performance. 
QoE is especially meaningful for applications that are latency and bandwidth sensitive, like video
streaming, where not all variations in path performance are easily absorbed, and some 
variations can result in behaviors (stalls, reduced quality) that are visible to the user.
QoE is helpful to researchers to characterize path suitability for applications~\cite{video_user_engagement, mobile_qoe, cfa, 5g_variegated}.
While QoE is difficult for 
network providers
to measure because it requires end system access and either running (or modeling) application behavior, 
that difficulty does not preclude an interest by network providers in QoE and methods to improve it~\cite{Mangla2019:Using,video_watching, liu_video}. 

Characterizing   cellular network performance is far more complicated. Cellular networks may
exhibit high performance variability over geographic space and time, affecting
a  user during application use. The physical (PHY) layer and radio access network (RAN) directly impact end-to-end performance. For instance, spectrum frequency, terrain and building materials all introduce complexities in signal propagation, available capacity, and even coverage stability and reliability~\cite{sonntag2013mobile, Zhou2004:Impact, 5g_variegated, 5g_mobility, yang2022mobile}.\footnote{As one example, 5G mmWave networks  can have greater performance instability than earlier versions of 5G due to the short, easily obstructable wavelengths~\cite{varshika_pam, mmwave_comparison, mmwave_coverage, ramadan_sigcomm24, ramadan_sigcomm2, Hassan2020:Channel}.}  Congestion in the RAN
caused by an overload of users competing for base station and PHY resources can produce
performance variability~\cite{Adarsh:IMC19, Adarsh:TMC21, cellular_crowds}. 
These factors are all inherent in cellular broadband, while none of them are
significant in fixed broadband. Together, these properties make cellular coverage fundamentally different from fixed broadband: performance is not merely variable but intermittently usable, rendering traditional QoS summaries and application-specific QoE metrics insufficient for characterizing coverage quality over space and time.

The difficulty of characterizing  cellular networks is evident in claims made by providers to both to the Federal Communications Commission (FCC) and the public.  Cellular coverage representations released by
the FCC in the National Broadband Map (NBM)~\cite{FCCNBB} indicate only whether a location is served at or above  a minimum speed (e.g. 35/3~Mbps download/upload speed for 5G) by a given  provider. Such
coarse characterizations of cellular service are not only insufficient~\cite{Adarsh:ICCCN21,Mangla:CACM22,challengesRWA,gaoReport} but are also frequently inaccurate~\cite{coverage_inaccurate1, coverage_inaccurate2, varshika_tprc}.  Public-facing claims  by cellular providers, although sometimes relying on third-party measurements~\cite{ookla_main,rootmetrics,jdpower,tmobile_jdpower}, are similarly vague and sometimes conflicting.\footnote{For example, 
in summer 2025, T-Mobile claimed to operate  ''America’s best network,'' citing Ookla’s performance data~\cite{ookla_report}.
    In October 2025, AT\&T launched a new advertising campaign accusing T-Mobile of ``untruths'' while promoting its larger, more reliable network~\cite{att_ad}.
     A September press release declared “Verizon remains America’s Best Wireless Network, Setting the Highest Bar for Quality and Reliability,”\cite{verizon_sep25}, citing results from drive-testing firm RootMetrics, which is owned by Ookla.}

\ignore{
The FCC is also interested in characterizing the performance of mobile broadband 
providers, in part due to a Congressional requirement to create and share
mobile broadband maps that aid consumers in decision making and holding providers accountable. 
Characterizations
of mobile broadband access released by the Federal Communications Commission (FCC) in the National Broadband Map~\cite{FCCNBB},  indicate whether a location is provided a minimum speed (e.g. 35~Mbps download/3~Mbps upload speed for 5G) by a particular cellular provider. 
While this provides slightly more information than the historical binary
representation of cellular coverage (i.e., does a location have access from a specific provider-yes or no), such
a simplistic characterization of networked services over cellular networks is insufficient ~\cite{Adarsh:ICCCN21,Mangla:CACM22,challengesRWA,gaoReport}. 
}

In this landscape of government activity, marketing claims, and third-party proprietary measurement, {\em our goal  is to  step back and
examine, from a research and scientific perspective, what metrics can   meaningfully capture the complexity of cellular network performance. } 
To this end, we introduce {\bf Quality of Coverage (QoC)},  a novel 
framework for more completely characterizing cellular network quality that emphasizes
usability and performance stability  over space and time. 
Similar to the definition of QoS and QoE, we define QoC as a set of key performance indicators (KPIs) that reflect network performance.
However, in defining QoC, we pay careful attention to the unique characteristics of the cellular  medium. 
To this end, we define five KPIs to characterize the quality and stability of a cellular network: \texttt{Usability}, \texttt{Usable Performance}, \texttt{Variability}, \texttt{Persistence}, and \texttt{Resilience}.  
Crucially, we do not need to measure each  KPI independently; each KPI is a function of 
network performance (e.g., download speed, latency) at a location over a period time. 


We offer the QoC framework as a robust starting point that we do not claim is, nor do we want to be, immutable. Similar to the evolution of QoS and QoE with the development of network and application-layer technologies, we anticipate that QoC, too, will evolve.    Our  goal is to   enable network researchers and practitioners to more meaningfully characterize cellular network performance.
By addressing the shortcomings of existing  evaluation methods and introducing a multi-dimensional framework for characterizing QoC, this paper sets a new standard in the field of cellular network characterization and assessment.  

In summary, our contributions  are the following:
\begin{enumerate}[leftmargin=*]
  \item \textbf{Definition of QoC}: We formalize QoC as a multi~‑~dimensional framework to quantify quality, usability and stability of cellular network performance.  We define five KPIs as a starting point for this framework. 
    \item \textbf{Spatial aggregation of  QoC profiles:}
Using distributional summaries, we show that spatial aggregation of QoC profiles preserves meaningful regional characterization while exposing heterogeneity typically hidden by current coverage representations.
    

    \item {\bf Multi-step characterization:}
We analyze the ability of  QoC KPIs to characterize a broad spectrum of network behaviors and examine KPI interdependence using both synthetic and real-world measurements.  
We demonstrate how QoC provides fine-grained characterizations of cellular performance that are impossible with current cellular benchmarking standards.
\end{enumerate}

By accounting for spatial and temporal fluctuations in coverage and performance, QoC   provides actionable insight for improving network   design, management, and 
 user experience. 
It can  enable network planners and local governments to
make  informed decisions about where new infrastructure is needed, and 
can help users identify  which provider offers the best local service. 
As new, hyper-local cellular technologies, such as mmWave 5G and 6G, are designed to support throughput- and latency-sensitive applications, QoC  offers a benchmarking framework tailored to environments where performance is inherently dynamic.

\noindent{\bf Ethical considerations.} This work does not use any user data.  All datasets are either synthetic or collected through author experimentation on research devices.



\vspace*{-0.1in}
\section{The Need for QoC}
\label{sec:motivation}
We begin by illustrating the lack of cellular-specific detail, and therefore inadequacy, of 
the minimum download speed representation of cellular coverage currently used by the FCC's NBM. 
We examine the coverage offered by AT\&T, T-Mobile and Verizon at 10 locations within our  community, covering roughly 20~sq. miles.  Based on both the NBM and each provider's coverage map, all three providers  offer download speeds  at a minimum of 35~Mbps (5G) at each  location.  
To analyze network coverage at these locations, we collected $\sim$4,200 Ookla Speedtest measurements using nine phones, three each of Samsung Galaxy S20+, S23, and Google Pixel 7 models, with identical cellular plans. At each location, we assigned one phone of each model to each carrier. 


\begin{figure*}[t!]
  \centering
   \begin{subfigure}[t]{0.31\textwidth}
   \centering
     \includegraphics[width=.9\textwidth]{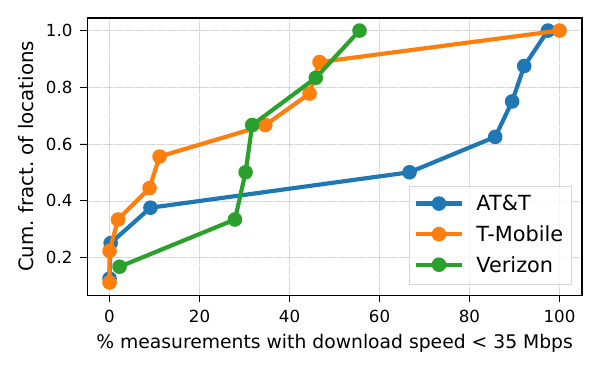}
    \caption{Sub-5G download speeds.}
     \label{fig:timeseries}
   \end{subfigure}
  \begin{subfigure}[t]{0.28\textwidth}
    \includegraphics[width=\textwidth]{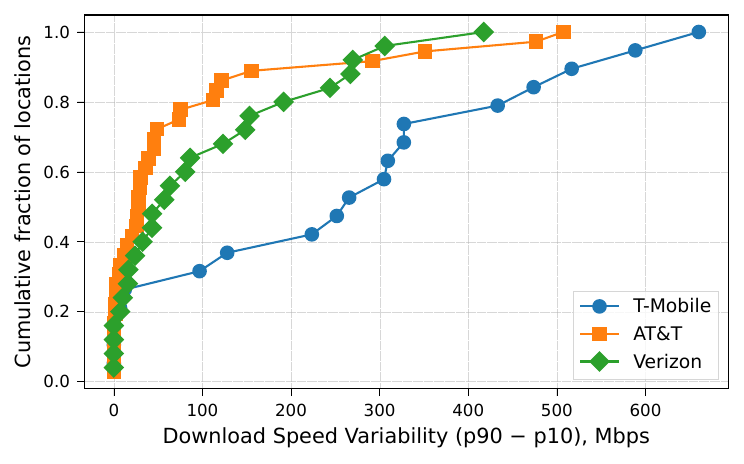}
    \caption{Intra-location speed variability.}
    \label{fig:variability}
  \end{subfigure}
  \begin{subfigure}[t]{0.30\textwidth}
    \centering
    \includegraphics[width=\textwidth]{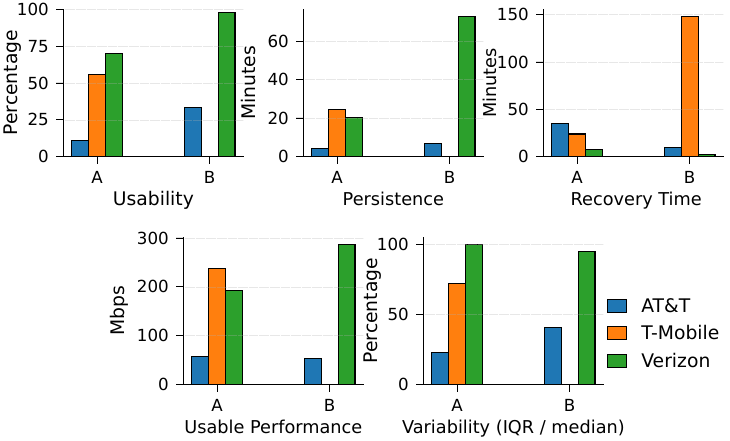}
    \caption{QoC KPIs for  2 example locations.}
    \label{fig:ookla_eval_1}
  \end{subfigure}
  \caption{Cellular performance at 52 locations in one community.}
  \label{fig:main}
\end{figure*}

The Speedtest measurements 
reveal a markedly different coverage picture than FCC and provider  maps. 
Per-location download speeds  range from 0.1 to 900~Mbps.  
AT\&T and Verizon average download speeds  measured less than  35~Mbps in  $\sim$45\% of  locations; nearly 17\% of  T-Mobile locations exhibit the same behavior.     
Figure~\ref{fig:timeseries} shows, for each location, the fraction  of measurements below the  35~Mbps threshold, revealing substantial  inter- and intra-location variability. While  some locations consistently exceed 35~Mbps,   50\% of AT\&T locations have more than 70\% of measurements below this threshold;   70\% of  locations have more than 30\% of Verizon's measurements failing it.

While this threshold-based view highlights widespread performance shortfalls, it does not capture the full extent of variability experienced at individual locations. To examine this variability more directly, we analyze the range of download speeds per location, which would be small if coverage quality were consistent.  Figure~\ref{fig:variability} shows this difference; each point represents the difference between the $10^{th}$ and $90^{th}$ percentile download speed at each location.  The results indicate that {\em all three providers exhibit significant performance variability  at the majority  of locations}. Notably  T-Mobile, despite offering the highest overall download speeds,  also shows the greatest variability: 50\% of   locations experience differences of 300 to 660~Mbps in download speeds. 


Together, these measurements highlight the challenge of representing cellular coverage quality.  Speed thresholds are often inaccurate depictions of  attainable bandwidth, and  attainable bandwidth can vary widely and unpredictability over space and time.  
We observe similar inter- and intra-location variability in RTT. 
Across all three carriers, 20\% of locations had $90^{th}$ percentile RTT values at least 100~ms above the median RTT.
Such instability and inconsistency in both bandwidth and latency can significantly affect the network’s ability to support throughput- and latency-sensitive applications at any given moment.

We conclude this case study by  demonstrating the fine-grained characterization ability of QoC. 
The QoC KPIs, defined in Section~\ref{sec:temporal}, characterize not just whether performance meets  a threshold, but how often that threshold is met and the consistency 
of that performance. 
To illustrate, Figure~\ref{fig:ookla_eval_1} shows the QoC KPIs for two example locations. 
At Location A, T-Mobile and Verizon offer comparable speeds (Usable Performance), yet their fine-grained performance differs. T-Mobile meets the 35~Mbps threshold 15\% less frequently than Verizon (Usability), 
but it  has 25\% lower Variability (IQR/median) than Verizon. At Location B, Verizon shows 100\% Variability in usable speeds, whereas AT\&T is more stable (IQR/median = 40\%). However, AT\&T has Usability of only 33\%, and the average length of continuous usable durations (Persistence) is only 6 minutes. Finally, T-Mobile never performs at the broadband threshold at this location.

These distinctions are critical for applications that require sustained connectivity, yet is absent from current coverage representations. Such characterizations are valuable both for consumers choosing between carriers and for researchers and broadband policymakers to better understand cellular coverage quality.


\section{Quality of Coverage Framework}


\subsection{Preliminary Insight}
\label{sec:Prelimin}

The case study in Section~\ref{sec:motivation} and prior work (e.g.,~\cite{reliability_nationwide,cellular_crowds}) demonstrate that cellular performance variability can be extreme, intermittently rendering the network unusable. We define \emph{usable} performance as that sufficient to complete common user tasks (e.g., loading a web page, sending a message, or joining a video call) without interruption. This motivates the QoC KPIs, which are designed to answer the following questions: (i) \emph{When is the network usable?} (ii) \emph{What performance level is typical during usable periods?} (iii) \emph{How much does performance vary when the network is usable?} (iv) \emph{How long do usable periods last?} and (v) \emph{How quickly does the network recover from performance drops?}

The KPIs we define in Section~\ref{sec:temporal} are designed to answer each of the above questions in turn. To do so, the foundation of the QoC framework is to first determine whether measured performance exceeds a predefined {\bf Usability Threshold} $\tau$. We instantiate $\tau$ as a benchmark reflecting the minimum performance requirements of the user tasks of interest, based on observed metrics such as throughput or latency; observations at or above\footnote{Performance at or above $\tau$ applies to metrics such as download or upload speed, while performance at or below $\tau$ applies to metrics such as latency. For simplicity, we describe $\tau$ in the context of the former, but note that the latter is also valid.} $\tau$ are labeled usable, and those below $\tau$ unusable.\footnote{This aligns with existing policy and measurement practices: the FCC~\cite{FCCNBB} defines 5/1~Mbps as the broadband benchmark for 4G and 35/3~Mbps for 5G-NR. While the FCC does not specify a latency benchmark for cellular networks, it requires that at least 95\% of latency measurements remain at or below 100~ms RTT for carriers receiving fixed broadband funding~\cite{usac_latency}.} Using this threshold, we label measurement windows as usable or unusable and compute QoC KPIs accordingly.

Understanding coverage quality, however, requires more than simply identifying whether a threshold is ever met. Early studies of cellular reliability and stability~\cite{Baltrunas2014:Reliability,bischof2017characterizingimprovingreliabilitybroadband} primarily relied on availability and percentile-based summaries, which characterize performance extremes extremes and instances of unreachability. Cellular coverage quality varies continuously over space and time, yet is observed in practice as discrete, timestamped measurements forming a time series. QoC operates directly on these measurements and the Usability Threshold $\tau$ to capture the combined impact  of cellular-specific factors, such as dynamic cell load, radio propagation, and scheduling. Capturing each of these influences individually is impractical, but their aggregate effect manifests as performance fluctuations in the time series data.

With $\tau$ defined, the performance time series naturally decomposes into a binary temporal process indicating usability at each time instant, along with two conditional performance distributions corresponding to usable and unusable periods. This separation distinguishes QoC from metrics based on overall performance, which can mask periods of unusable performance by averaging them with high-performing samples. Instead, QoC focuses explicitly on intervals where performance is above $\tau$, characterizing both typical performance and its variability during usable periods to better reflect user experience.

\begin{figure}[t]
  \centering
    \vspace{-\intextsep}
  \includegraphics[width=0.35\textwidth]{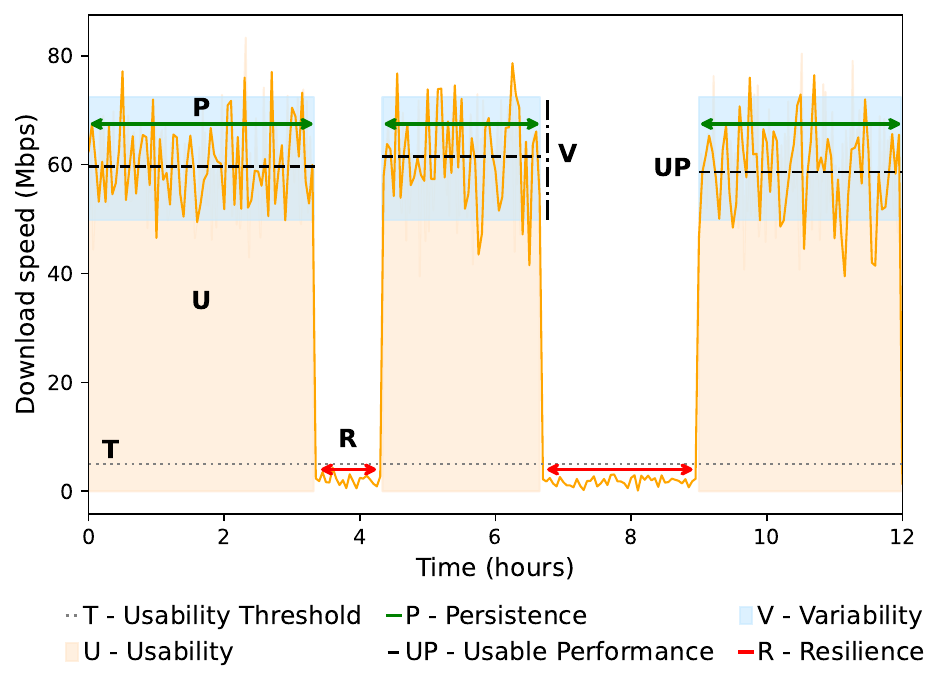}
  \caption{Illustration of QoC KPIs. Green arrows denote usable periods $L_i \in \mathcal{L}$; red arrows denote unusable periods $D_j \in \mathcal{D}$.}
  \label{fig:qoc_illus}
   \vspace{-0.1in}
\end{figure}

\subsection{QoC Key Performance Indicators}
\label{sec:temporal}

In the following definitions, at each geolocation $(x,y)$, let $X(x,y,t)$ denote the network performance measurement (e.g., download speed or latency) 
at time $t$, such that all KPIs  are computed for the specific location $(x,y)$. Formally, all KPIs are defined in continuous time for generality, but in practice, measurements are obtained as discrete samples over finite time intervals; thus, integrals are realized as summations over $T$ sampled time points.
For reference, Figure~\ref{fig:qoc_illus} shows how each  component of the QoC profile $\mathbf{Q}(x,y)$ is derived from an illustrative 12-hour time series of download speed measurements. Table~\ref{tab:qoc_properties} in the Appendix summarizes  the expected behavior of each QoC KPI as $\tau$ varies.
\ignore{
To emphasize observed performance quality, \texttt{Usability} serves as the basis for all  QoC KPIs. 

Hence, a meaningful assessment of QoC must also capture how stable network performance is over time. To this end, we introduce two complementary dimensions of stability: \texttt{Persistence} and \texttt{Resilience}.
We define \texttt{Persistence} as the average duration for which performance remains continuously above the \texttt{Usability Threshold}. Inspired from Paxson’s path-lifetime constructs~\cite{Paxson1997:End} and complementing conventional metrics, such as mean time between failures, we focus on the longevity of usable service rather than the frequency of disruptions. \texttt{Resilience}, in contrast, captures how quickly performance recovers after falling below the \texttt{Usability Threshold}. 
Together, \texttt{Persistence} and \texttt{Resilience} characterize  temporal stability of location-specific coverage beyond what point-based snapshots can provide. 
}

\vspace*{0.1in}
\noindent
{\textbf{Usability (U)}:} 
\texttt{Usability} measures the fraction of time the network delivers acceptable performance; we term these "usable periods." 
Our goal is not to select an appropriate threshold, but rather to provide a framework for defining a \texttt{Usability} KPI. Let $\tau$ be the predefined Usability Threshold for metric $X$ (e.g. 100~ms for latency or 35~Mbps for download speed). 
Over a continuous observation window of duration $T$, \texttt{Usability} is defined as:
\vspace{-0.1in}
\begin{equation*}
U = \tfrac{1}{T}\int_{}^{T} \mathbb{I}\big(\mathsf{usable}_\tau[X(x,y,t)]\big)\,dt,
\end{equation*}
where $\mathsf{usable}_\tau[\cdot]$ denotes the predicate that evaluates to $1$ when the performance metric satisfies the usability condition (i.e., $X(x,y,t)\!\ge\!\tau$ for uplink/downlink bandwidth). 
We denote the measurements $X(x,y,t)$ for which $\mathsf{usable}_\tau[X(x,y,t)]$ holds as $X_{\text{usable}}(x,y,t)$.
The \texttt{Usability} KPI serves as the basis for all  QoC KPIs. 


\noindent
{\textbf{Usable Performance (UP):}}
While Usability  provides a baseline,  \texttt{Usable Performance} 
quantifies user-experienced  performance during all usability periods.
Over a continuous observation window of duration $T$, Usable Performance is characterized by the distribution of values observed during usable periods. If a single scalar is desired, it can be any chosen summary of $X_{\text{usable}}(x,y,t)$, e.g., a percentile or mean, as:
\vspace{-0.1in}
 \begin{equation*}
UP_{\text{usable}}
=
\tfrac{\int_{}^{T} X_{\text{usable}}(x,y,t)\,dt}
     {\int_{}^{T} \mathbb{I}\!\left(\mathsf{usable}_\tau[X(x,y,t)]\right)\,dt}
\end{equation*}
 \vspace{-0.1in}
 

\vspace*{0.1in}
\noindent
{\textbf{Usable Performance Variability (V)}:}  
While all measurements within a usable period satisfy the usability condition, the performance may  fluctuate. Since our focus is on cellular coverage quality, we are interested in capturing the spread of performance during usable periods. 
There are many ways to define  performance variability. We use the interquartile range (IQR)/median because it is robust to outliers and well-suited for the skewed distributions typical of network performance data, providing a more stable measure of variability.
Over a continuous observation window of duration $T$, \texttt{Usable Performance Variability} is defined as:
\vspace{-0.1in}
\begin{equation*}
V_{\text{usable}}
=
\tfrac{
P_{75}\!\Big(\{X_{\text{usable}}(x,y,t)\}\Big)
-
P_{25}\!\Big(\{X_{\text{usable}}(x,y,t)\}\Big)
}{
P_{50}\!\Big(\{X_{\text{usable}}(x,y,t)\}\Big)
}, \ t\in[0,T]
\end{equation*}
where $P_{75}$, $P_{50}$ and $P_{25}$ are the 75$^{th}$, 50$^{th}$ and 25$^{th}$ percentiles, respectively, of performance over all $X(x,y,t)$ such that $\mathsf{usable}_\tau[X(x,y,t)]$ holds.


\vspace*{0.1in}
\noindent
{\textbf{Persistence (P):}}
A meaningful assessment of QoC must  capture how stable network performance is over time. To this end, we introduce two complementary dimensions of stability.  The first is \texttt{Persistence}, defined using the distribution of durations for which performance remains continuously above $\tau$. Inspired from Paxson’s path-lifetime constructs~\cite{Paxson1997:End} and complementing conventional metrics, 
such as mean time to failure (MTTF)~\cite{reliability_metrics}, 
\texttt{Persistence} reflects temporal stability by quantifying the 
duration of continuous usable periods. 
Suppose there are $N$ continuous usable periods
during which $\mathsf{usable}_\tau[X(x,y,t)]$ holds. Let $\{L_j\}_{j=1}^{N}$ denote the distribution of usable-period durations, where $L_j$ is the duration of the $j^{th}$ period, beginning at time $t_j^{start}$ and ending at time $t_j^{end}$, and defined as $
L_j = \int_{t_j^{\mathrm{start}}}^{t_j^{\mathrm{end}}} \mathbb{I}\big(\mathsf{usable}_\tau[X(x,y,t)]\big)\,dt.$

\noindent
If a single scalar is desired, it can be any chosen summary of $\{L_j\}$, e.g., a percentile, or the mean, as:
\vspace{-0.1in}
\begin{equation*}
P_{\text{usable}} = \tfrac{1}{N}\sum_{j=1}^{N}L_j
\end{equation*}
 \vspace{-0.1in}

\noindent
{\textbf{Resilience (R)}:}  The second dimension of stability is \texttt{Resilience}, which quantifies the network’s ability to recover from performance drops below $\tau$, and is defined in terms of mean time to recovery (MTTR)~\cite{reliability_metrics}. 
Let there be \(W\) continuous unusable periods during which  $\mathsf{usable}_\tau[X(x,y,t)]$ does not hold,
 and let $\{D_i\}_{i=1}^{W}$ denote the distribution of unusable-period lengths, where $D_i$ is the duration of the \(i\)$^{th}$ unusable period, beginning at time $t_i^{start}$ and ending at time $t_i^{end}$ defined as 
$D_i = \int_{t_i^{\mathrm{start}}}^{t_i^{\mathrm{end}}} \big(1 - \mathbb{I}(\mathsf{usable}_\tau[X(x,y,t)])\big)\,dt$; this is the {\em recovery time}.  \texttt{Resilience} is defined as the inverse of the 
recovery time. This can be any chosen summary of $\{D_i\}_{i=1}^{W}$, e.g., a percentile, or the mean, as shown below:
 \vspace{-0.1in}
\[
R = \tfrac{W}{\sum_{i=1}^{W} D_i}\,.
\]
\noindent
{\textbf{Complete  QoC Profile}: }
At each location $(x,y)$, we define the  QoC as the five‐dimensional profile:
\vspace*{-0.05in}
\begin{equation}
\mathbf{Q}(x,y) \;=\;\bigl(
  U,\;\!UP_{usable}, V_{usable},\; P_{usable},\;R
\bigr).
\label{eq:qoc_vector}
\end{equation}
\vspace*{-0.15in}
\noindent

\noindent
By incorporating  traditional usability and performance metrics along with  stability representations, our framework enables a richer characterization of cellular coverage that is more responsive to changes in performance over time.

\subsection{Spatial Aggregation of QoC Profiles}
\label{sec:spatial}

We now consider how QoC of individual spatial units can be 
aggregated over larger regions. We are motivated by  questions city planners,  residents, or network researchers may ask, such as: \textit{What coverage quality does each  provider offer within a community?} \textit{Which regions have extended periods of poor network performance, indicating they need  upgrades?} and \textit{How homogeneous is the coverage quality  within a census tract?} 
To answer them, we can aggregate the QoC KPIs at individual units into meaningful region-level summaries.
Here, we use “region” to mean any spatial unit of interest:   a census tract, neighborhood, uniform H3 hexagon (see Section~\ref{sec:synthetic}), or custom polygon.  We  apply the same aggregation logic to whichever unit best suits policy goals.  


Cellular networks  exhibit \emph{spatial heterogeneity}; base station density, interference patterns, and user load change--potentially dramatically--between  regions. Hence, to extend the QoC framework from an individual spatial unit 
to regions, we employ a 
two-step aggregation strategy via  scalar averages and distribution-preservation summaries. This strategy  preserves the richness of individual behavior yet offers the efficiency required for large-scale spatial analysis.

Each individual spatial unit
is associated with a
QoC profile derived from its network measurement time series. To construct an aggregate QoC profile, 
we (1)~retain the \emph{entire distribution} of each QoC KPI across the unit level and then (2) aggregate these distributions to form region-level QoC summaries.
This 2-step methodology offers key advantages:
\setlist{nolistsep}
\begin{itemize}[leftmargin=*]
     \item \textbf{Heterogeneity preservation:} 
     Region-level distributional summaries capture the spread and tails of each QoC KPI, ensuring that localized weak-coverage pockets are not obscured by better-performing areas.
     \item \textbf{Fine-grained quantile queries:} Retaining distributions enables estimation of various quantiles of the distribution of each QoC KPI (e.g., 10$^{th}$, 50$^{th}$, 90$^{th}$ percentiles), supporting queries such as "What is the 10$^{th}$ percentile of \texttt{Resilience} across the region?"~\footnote{When storing entire distributions is impractical, one can store the distributional information via compact distribution-preserving summaries, such as quantile sketches~\cite{datadog_sketch, kll_apache, masson2019ddsketch}, that support approximate percentile queries and hierarchical mergeability: summaries from smaller regions can be efficiently merged into larger-region summaries.}
\end{itemize}
In addition to the distribution summaries, we compute the average of each QoC KPI across all constituent spatial units. For each QoC
KPI $m$, the region-level average  
    $\overline{m}_{\text{region}} = \tfrac{1}{M} \sum_{i=1}^{M} m_i$, 
where $M$ is the number of units in the region and $m_i$ is the average per-unit QoC KPI value. This average vector complements the percentile-based distribution summaries by offering a representation of typical spatial QoC. 
This two-fold approach provides flexibility in answering a variety of question types involving larger regions, and
ensures that both distributional summaries and average trends are accessible without recomputation. 
Finally, both the distributional and average-based aggregation methods preserve the five-dimensional structure of the QoC representation. Each region is summarized as the aggregated QoC profile for each constituent spatial component. 


\vspace*{-0.02in}
\section{QoC Characterization Datasets}

\begin{table*}[t]
\centering\scriptsize
  \vspace{-\intextsep}
\setlength{\tabcolsep}{3pt}
\begin{tabular}{@{}llp{4.2cm}p{5.5cm}@{}}
\toprule
{\bf Scenario} & {\bf Distribution Model} & {\bf Expected Behavior} & {\bf Download Speed Selection Parameters} \\ 
\midrule
Persistent-good (PG)  
& Normal 
& Consistently high and stable performance.
& $\mu=500\pm5\%$, bounds:[400,600], $\mathrm{CV}=0.05$ \\ \hline
Persistent-poor (PP) 
& Normal 
& Consistently low and stable performance. 
& $\mu=5\pm5\%$, bounds:[1,20], $\mathrm{CV}=0.05$ \\ \hline
Periodic
& Time-varying Mean 
& Sinusoidally modulated time-varying mean to show diurnal fluctuations.
& $\mu(t)=\mu_{\mathrm{base}}+A\cos(4\pi t/1440)$, $\mu=500\pm1\%$, $A=500\pm1\%$, bounds:[1,1000] \\ \hline
Variable 
& Log-Normal 
& Right-skewed variability typical of real network traffic~\cite{Alasmar2021:LogNormal}
& $\mu=\ln(500)\pm5\%$, $\sigma=2.5$, bounds:[1,1000] \\ \hline
Short-frequent drops (SFD)  
& Hidden Markov Model (N) 
& Short, but frequent disruptions ($\sim$30\,min down / $\sim$3\,hr up); 15\% unusable
& Usable: $\mu=500\pm5\%$, [400,600], $\mathrm{CV}=0.05$; 
  Unusable: $\mu=2\pm5\%$, [1,5], $\mathrm{CV}=0.20$ \\ \hline
Long-rare drops (LRD) 
& Hidden Markov Model (N) 
& Infrequent, but prolonged disruptions ($\sim$12\,hr down / $\sim$3\,days up); 15\% unusable
& Usable: $\mu=500\pm5\%$, [400,600], $\mathrm{CV}=0.05$; 
  Unusable: $\mu=2\pm5\%$, [1,5], $\mathrm{CV}=0.20$ \\ \hline
Congestion 
& Hidden Markov Model (N)  
& Chronic slowdowns with occasional relief ($\sim$6\,hr congested / $\sim$1\,hr relief)
& Congested: $\mu=10\pm5\%$, [5,25], $\mathrm{CV}=0.20$;  
  Relief: $\mu=50\pm5\%$, [30,50], $\mathrm{CV}=0.30$ \\ \hline
\bottomrule
\end{tabular}
\caption{Summary of simulated network scenarios with compact parameter notation.}
\vspace*{-0.3in}
\label{tab:qoc-scenarios}
\end{table*}

There are two primary goals for our evaluation. First, we assess the ability of the QoC KPIs to characterize network performance at both individual locations and aggregated geographic regions, with an emphasis on capturing fine-grained performance variability. This requires datasets that exhibit diverse network behaviors. Second, we evaluate the sensitivity of the QoC KPIs to temporal and spatial measurement density, motivated by real-world constraints that often limit the frequency and coverage of cellular measurements. Hence, our evaluation uses temporally and spatially dense data that can be systematically down-sampled to study density sensitivity.

To this end, we use two novel datasets.\footnote{Upon acceptance, both datasets will be publicly released.} The first is collected from AT\&T’s production network within our community. Both the FCC NBM and AT\&T’s coverage map report 5G coverage with a minimum service level of 35/3~Mbps download/upload throughout the area. As shown in Section~\ref{sec:motivation} and confirmed by additional analysis, AT\&T exhibits the highest performance variability among major providers locally. We therefore focus on AT\&T, as it provides an ideal setting for evaluating QoC’s representation of stability, reliability, and fine-grained performance dynamics.  Second, we generate a synthetic dataset with seven predefined network behaviors, enabling controlled evaluation of QoC’s ability to characterize expected performance patterns.

\noindent
{\bf AT\&T Network Measurement Data.}
\label{sec:realworld}
We procured 15 SIM cards from AT\&T subscribed to identical cellular plans with unlimited data capacity, and deployed one phone\footnote{Selected from: Samsung Galaxy S20+, S23, S25, and Google Pixel
7 and 9.}  at each of 15  locations across an 8 sq. mile local region. 
Our data was collected in January 2026; each location captured  seven  days of  measurements.  We developed an automated Android measurement setup that  (i) issues UDP probes to our university-hosted echo server and reads the cellular signal state every 500~ms, 
and (ii) runs Ookla Speedtest measurements every 10~minutes. 
RTT and signal information provide continuous visibility into short-lived degradations and stability, while periodic Speedtests provide  application-level throughput. Together, these measurements enable us to characterize  a location’s performance over time, as reflected by the QoC KPIs. 
Further, the data can be aggregated to study the spatial representation of QoC across this region. 
In total, the dataset comprises $\sim$15~million RTT measurements and $\sim$15,000 Speedtests. 

\enlargethispage{\baselineskip}
\noindent
{\bf Synthetic Datasets.}
\label{sec:synthetic}
To analyze the accuracy and characterization ability of QoC, an understanding of expected behavior is critical.  To this end, we generated a detailed synthetic dataset to represent seven distinct,  predefined network scenarios.\footnote{We focus our synthetic data study on the KPI  representation of download speed due to the prominence of download speed as an indicator of network performance and because of its larger range of variability compared to  metrics like upload speed and latency.  However, our evaluation could easily be replicated for latency or other network metrics.} 
To organize the data spatially, we employ Uber’s~\cite{uber_h3} H3 hexagonal indexing system, which the FCC~\cite{FCCNBB}  uses to map broadband coverage.\footnote{\emph{H3} defines a hierarchy of spatial resolutions: at resolution-0 a single hexagon covers the entire Earth. Successive partitions produce increasingly finer hexagons up to resolution 15.  For example, resolution-8 ("hex8") cells average $0.73\ \mathrm{km}^2$, resolution-9 ("hex9") cells $0.10\ \mathrm{km}^2$, and resolution-10 ("hex10") cells $0.014\ \mathrm{km}^2$.  Each parent hexagon has seven child hexagons.} 
We utilize hex10 hexagons as the fundamental simulation unit; each generated data point represents performance within one hex10 hexagon. 
For each of the seven predefined scenarios, we generate \textit{seven} time series, one per \textit{hex10} spatial unit.  Each time series  represents download speed measurements  every minute for a 30-day period. This results in 43,200 measurements per time series and 302,400 measurements per scenario. 
To ensure statistical robustness, we run 50
Monte Carlo simulations for each scenario, 
yielding  $\sim$15~million measurements per scenario 
and $\sim$105~million measurements across all seven scenarios. 

To study QoC KPIs at an individual spatial unit, we are interested in  characterization of performance over time. We use one hex10 time series per scenario (and its 50 simulations) for analysis. 
To evaluate QoC spatial aggregation, we utilize all seven hex10 time series per scenario (and their 50 simulations) and aggregate them at the hex9 level. 
We summarize these measurement  details 
in Table~\ref{tab:datagen_breakdown} in the Appendix.

Table~\ref{tab:qoc-scenarios} summarizes the statistical parameters used by each scenario and its expected behavior.  The Persistent-good and Persistent-poor scenarios model consistent performance, while Periodic introduces regular fluctuations with controllable variance.  The Variable scenario models right-skewed variability typical of real networks~\cite{Alasmar2021:LogNormal}.
To simulate extended periods of unusable performance for the short frequent drops (SFD), long rare drops (LRD), and congestion scenarios, we use a two-state Hidden Markov Model (HMM).\footnote{An HMM is a stochastic process in which an unobserved (“hidden”) state variable evolves over time according to fixed transition probabilities, while observations are drawn from a state-dependent distribution.} The hidden variable has two values: state 0 (usable) and state 1 (unusable). At each time step, the HMM selects  the network state.
The state transition matrix is defined as:
\begingroup
\vspace{-0.02in}
\setlength\arraycolsep{2pt}
  \scriptsize                  
  \[
 A = \bigl(\begin{smallmatrix}
1 - p_{\mathrm{entry}} & p_{\mathrm{entry}}\\[-2pt]
1 - p_{\mathrm{self}} & p_{\mathrm{self}}
\end{smallmatrix}\bigr),
\; p_{\mathrm{entry}} = P(s_t{=}1\mid s_{t-1}{=}0),
\; p_{\mathrm{self}} = P(s_t{=}1\mid s_{t-1}{=}1).
\]
\endgroup

\noindent
This model controls how often performance drops begin ($p_{\mathrm{entry}}$) and how long they persist ($p_{\mathrm{self}}$). Within each state, we simulate download speed values using a normal distribution. The usable and unusable states are described in Table~\ref{tab:qoc-scenarios}. 
In the SFD and LRD scenarios, the network is expected to be unusable for the same total duration (15\% of total time), but the frequency and duration of disruptions vary.

\begin{figure*}[t]
  \centering
    \begin{subfigure}[b]{0.42\linewidth}
    \centering
    \includegraphics[width=0.95\linewidth]{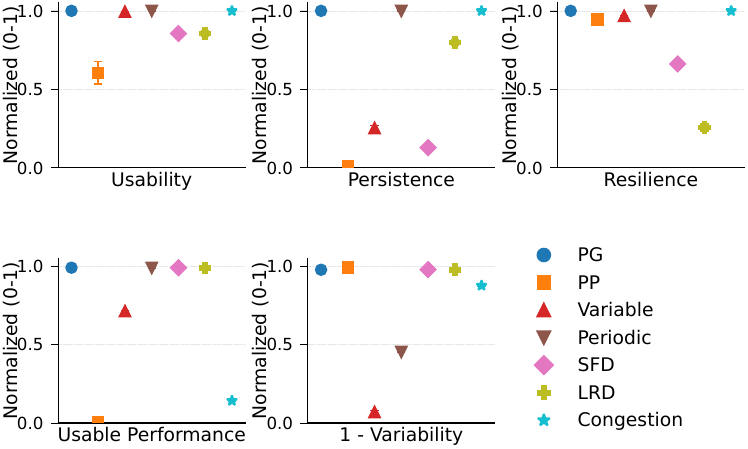}
   \caption{$\tau$ = 5 Mbps}
   \label{radar_5}
 \end{subfigure}
 \hspace{0.2in}
  \begin{subfigure}[b]{0.42\linewidth}
    \centering
    \includegraphics[width=0.95\linewidth]{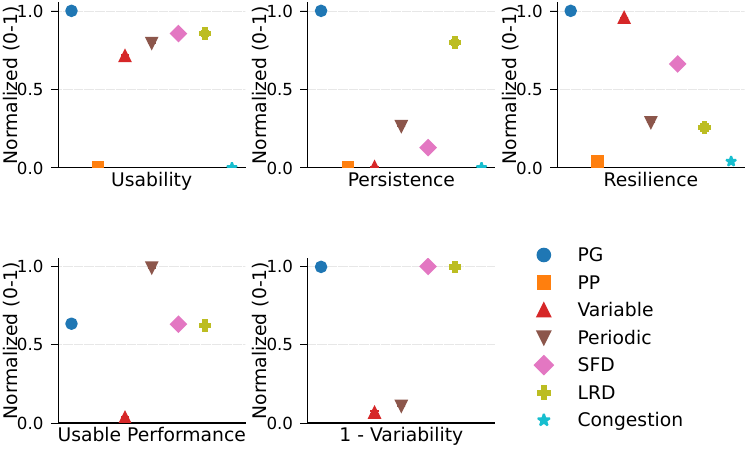}
    \caption{$\tau$ = 100 Mbps}
    \label{radar_100}
  \end{subfigure}
  \caption{QoC KPIs of synthetic scenarios.
  }
    \label{qoc_comparisons}
\end{figure*}

The synthetic scenarios are not intended to represent a complete set of  performance environments nor to   precisely reflect real  performance.  Our goal is a set of controlled, well-understood, diverse behaviors in which we can evaluate QoC KPI characterizations.   Figure~\ref{fig:syn_simulations}  in the Appendix shows the download speed values generated by each  scenario, and Table~\ref{tab:KS_gendata} shows the two-sample
Kolmogorov-Smirnov (KS) test statistic comparing these distributions, to demonstrate their statistical difference.

\section{QoC and Expected Behavior}
\label{sec:qoc_synthetic}

The first step in our evaluation of QoC is to verify its ability to accurately characterize cellular performance.  We are particularly interested in its ability to 
 distinguish fine-grained network behavior such as  temporal (in)stability at varying timescales.  Hence, 
in this section, we apply QoC to the seven synthetic  scenarios to evaluate  QoC representation of  performance timeseries with different temporal patterns.  
\ignore{
In Section~\ref{sec:parameters}, we analyze the sensitivity of QoC KPIs to the Usability Threshold and other parameters.  
We study the spatial aggregation properties, and in particularly the preservation of per-location stability differences, in Section~\ref{sec:aggregation}.  In our evaluation, we utilize different combinations of the synthetic and real-network datasets, in each case basing our analysis on the dataset(s) that can be explore the research goal.  Because the Usability Threshold is a benchmark can be applied to any performance requirement of interest, we demonstrate its application to both download speed and latency.
}

Each KPI 
is measured on different scales and with units; we retain the raw units to interpret coverage quality.
For visual clarity, we apply standard normalization techniques~\cite{min_max, log_transform}, using natural log transformation and min–max scaling to map all KPIs to a common [0,1] interval. This facilitates direct comparison and joint analysis. For consistency in interpretation between QoC KPIs, we ensure higher scores denote better performance. Therefore, we invert the normalized Variability KPI to (1 - V) and utilize this value in our plots.



\vspace*{-0.1in}
\enlargethispage{\baselineskip}
\subsection{Descriptive Ability of QoC}

We compute QoC KPIs on the synthetic scenarios for $\tau$ = 5~ and 100~Mbps, broadly representing  bandwidth requirements of different applications.\footnote{These thresholds  align with FCC definitions 
for coverage reporting and policy eligibility. For mobile networks, speeds below 5~Mbps qualify as "unserved", and  for fixed networks, speeds below 100~Mbps are "underserved".}
We  plot the means and 95\% confidence intervals
of each normalized KPI  in Figure~\ref{qoc_comparisons}.\footnote{Note the confidence intervals are very small and therefore not visible on most points.}  

\begin{figure}[t]
  \vspace{-\intextsep}
  \centering
    \centering
    \includegraphics[width=0.9\linewidth]{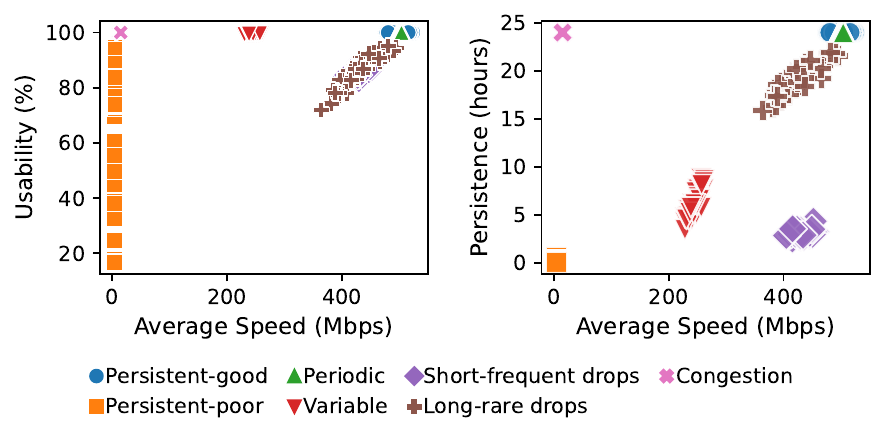}
  \caption{QoC KPI characterization ability  in synthetic scenarios. $\tau$= 5~Mbps.}
\label{fig:avg_comparison}
\vspace*{-0.1in}
\end{figure}

\begin{figure*}[t]
  \vspace{-\intextsep}
\centering
\begin{subfigure}[b]{0.3\linewidth}
\includegraphics[width=\textwidth]{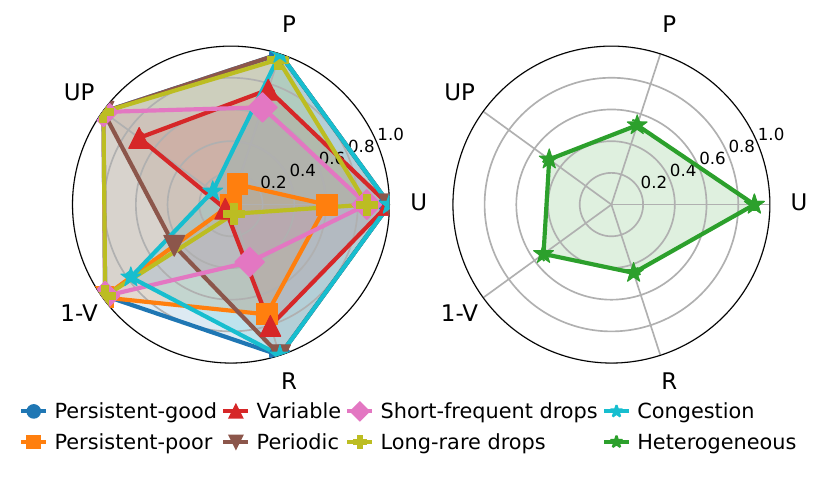}
\caption{Spatial QoC: homogeneous (left) and heterogeneous (right)}
\label{fig:spatial_radar}
\end{subfigure}\hspace*{0.3in}
\begin{subfigure}[b]{0.25\linewidth}
\includegraphics[width=\textwidth]{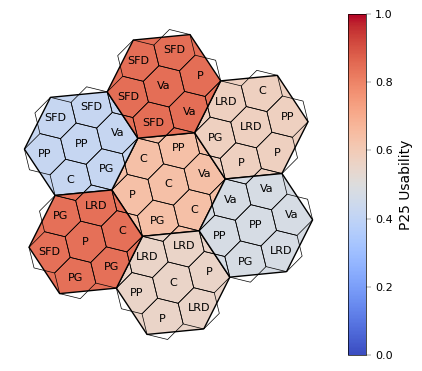}
\caption{$25^{th}$ percentile Usability \label{case1}} 
\end{subfigure}%
\begin{subfigure}[b]{0.25\linewidth}
\includegraphics[width=\textwidth]{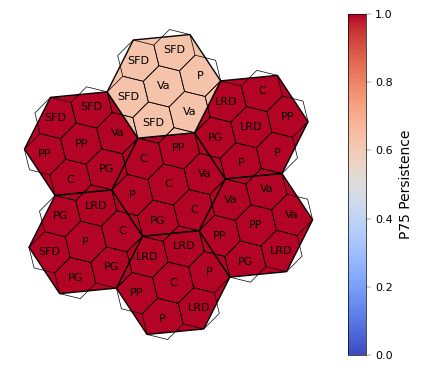}
\caption{$75^{th}$ percentile Persistence\label{case2}} 
\end{subfigure}%
\caption{Spatial QoC KPIs for homogeneous, heterogeneous and random assignments, and at $\tau$= 5~Mbps.} 
\label{fig:spatial_random}
\vspace*{-1em}
\end{figure*}

The  QoC KPIs reflect high Usability scores for almost all synthetic networks when $\tau$ = 5~Mbps, as shown in Figure~\ref{qoc_comparisons}(a).  This is expected because  most selected download speeds are well above $\tau$.  PP networks  are the notable exception, since their download speed mean is 5~Mbps. 
PP networks also have  lower  Persistence, Usable Performance and Variability, but high Resilience; Resilience is high because $\tau$ is equal to the mean download speed in this scenario, 
and so speeds frequently oscillate around this value. 
Periodic behavior networks  have similar KPIs to PG networks except for Variability. Despite the regular performance drops in Periodic networks, download speeds still stay above  5~Mbps; hence most KPIs demonstrate strong performance.  However, the lower Variability score indicates the download speeds are less consistent than the PG, and even PP, networks. 

The PG networks  score highly on all KPIs.
Networks with Variable performance (red triangle) are characterized by high Usability and Resilience due to the low value of $\tau$.
They also
show the highest Variability score.  The low Persistence score characterizes the bursty nature of the download speeds in this scenario.
In Congested networks, Usability, Persistence, and Resilience remain high. Though this may seem counterintuitive, it is because the 
$\tau$ = 5~Mbps is so slow; performance fluctuations rarely dip into unusable bandwidths, and so these KPIs remain high. However, the Usable Performance and Variability KPIs show lower values than the other scenarios, characterizing the lower download speeds and high variability during usable periods.

Finally, SFD and LRD network behaviors share similar Usability scores yet differ in Persistence and Resilience by about 20\%. These KPIs reveal the performance differentials in these two network types: frequent small interruptions (SFD) produce a lower Persistence KPI, while infrequent but prolonged periods of unusability (LRD) result in higher Persistence but lower Resilience because it takes longer to return to a download speed above  $\tau$. This distinction matters in practice: applications such as video streaming or VoIP may tolerate brief, infrequent drops but degrade significantly under frequent or prolonged disruptions, while access to emergency services (e.g. calling 911) may prioritize quick recoveries from disruptions over sustained "good" performance.

When the value of $\tau$ increases to 100~Mbps (Figure~\ref{radar_100}), QoC KPIs in each scenario adapt accordingly.  For example, the KPIs become zero for the Congested network because its performance never reaches 100~Mbps. Usability, Persistence and Resilience  drop in the Variable network since download speeds are frequently below 100 Mbps. 
Interestingly, Variability also decreases because the range of usable download speeds (> 100 Mbps) is smaller.

\noindent
{\bf Characterization of variable/unstable network performance.}
The power of the QoC framework is most apparent in its ability to distinguish 
network variability.
We  illustrate this in 
Figure~\ref{fig:avg_comparison}, which shows a scatterplot of the relationship between the daily average download speeds and the daily Usability (left) and  Persistence (right) with $\tau$ = 5~Mbps. PG, LRD, SFD, and Periodic networks (clustered in the top right corners)  exhibit high average speeds; yet they show distinct Usability and Persistence values.  Usability ranges from 65\% to 100\% and  Persistence ranges from five to 24~hours. In both cases, LRD networks  show the lowest values. 

As another example, networks with different stability behaviors (e.g., Periodic and Congestion) can exhibit identical Usability (100\%) when $\tau$=5~Mbps, yet different Usable Performance (shown in Figure~\ref{radar_5}).  Conversely, networks with higher Usable Performance can also have lower Usability values, as shown by LRD networks.\footnote{This implies these networks more often exhibit speeds below $\tau$, but when speeds are above $\tau$, they are high.} 
Figure~\ref{fig:drops} in the Appendix highlights the behavior of the SFD and LRD networks, which are both unusable for 15\% of the time.  The figure shows that while Usability is the same, Persistence and Resilience differ significantly.
Finally, we confirm that the KPIs capture different aspects of temporal stability through the low and varying levels of mutual information observed within the KPIs, as shown in Figure~\ref{fig:mut_info} of the Appendix.

\vspace*{-0.1in}
\subsection{QoC Spatial Aggregations}
\label{sec:synth_spatial}
We next evaluate the spatial aggregation of QoC to   demonstrate how  
performance data of individual spatial units can be aggregated into larger spatial regions (i.e., communities, census blocks). An important step is to determine whether differences in per-location stability are lost in the aggregation. 
As described in Section~\ref{sec:synthetic}, we model each spatial region as a hex9 hexagon, each of which consists of seven hex10 cells. We utilize all seven  generated  download speed time-series per each of the seven network scenarios (49 time-series in total) to compute spatial QoC. This yields $\binom{49}{7}$ possible combinations to cluster the 49 hex10s into groups of seven, each forming a hex9 region.
We construct two distinct hex9 spatial configurations: 
(i) \textit{homogeneous}: each hex9 region consists of seven hex10s from the same network scenario,
resulting in seven spatially homogeneous regions per simulation.  This combination enables study of whether the aggregation of QoC KPIs under uniform network scenarios preserves their underlying individual behaviors.
(ii) \textit{heterogeneous}: each hex9 contains one hex10 from each of the seven network scenarios. This creates seven balanced regions where each network scenario is equally represented and enables analysis  of QoC's  characterization of mixed spatial environments.

\noindent{\bf Preservation of per-location stability differences.}  
Figure~\ref{fig:spatial_radar} 
shows radar plots of the aggregated QoC profiles for each region at $\tau$= 5~Mbps. 
In the homogeneous aggregation (left plot),  the average of each normalized QoC KPI in all   network scenarios behaves identically to the individual  QoC KPIs in that network scenario that were shown in Figure~\ref{qoc_comparisons}. 
In the heterogeneous aggregation (right plot), 
the QoC KPIs show more moderate values due to the controlled  mix  of each network scenario per hex9.

To evaluate whether spatial aggregation preserves meaningful differences in per-location stability, we illustrate two examples. 
We construct a random assignment of the 49 hex10s grouped into seven hex9 regions, such that they are neither homogeneous nor heterogeneous; Figures~\ref{case1} and ~\ref{case2}
show the spatial layout.  We annotate each hex10 with its assigned network scenario. Using the distributional information, we estimate the $25^{th}$ percentile {Usability} for each hex9 region in Figure~\ref{case1}, and the $75^{th}$ percentile Persistence for each hex9 region in Figure~\ref{case2}.

The aggregated KPIs confirm that per-location stability differences are preserved through spatial aggregation. The $25^{th}$ percentile {Usability} is highest in hex9 regions dominated by hex10 regions with high Usability: that is, the PG, Va, SFD, and LRD networks. As the number of PP or Congested (C) hex10s in a hex9 increases, $25^{th}$ percentile {Usability} decreases, demonstrating that the instability characteristics of individual locations propagate meaningfully to the regional level.
Since the $\tau$ is 
low, the $75^{th}$ percentile Persistence does not drop across any hex9 region. However, when dominated by SFD networks, the 75th percentile Persistence decreases noticeably due to the frequent short interruptions characteristic of this network scenario.
These results confirm that spatial aggregation retains the distinguishing stability patterns of constituent hex10 locations.



\vspace*{-0.1in}
\section{QoC on Production Networks}
\label{sec:production}

We turn our focus to the application and evaluation of   QoC  to AT\&T's production network to understand its utility in a live network. 
Using the AT\&T  measurements from the 15 locations described in Section~\ref{sec:realworld}, we compute the QoC KPIs for {\em both} download speed and RTT Usability Thresholds to determine what QoC reveals about cellular coverage in our community.\footnote{Recall that the FCC National Broadband Map and the AT\&T coverage map indicate all measurement locations have 5G 35/3~Mbps download/upload speeds. They do not provide any additional details about performance.} 
Because the locations are labeled as having 35/3~Mbps 5G coverage, we use this threshold as  $\tau$ to determine whether download/upload speeds reveal  usable performance. Though the FCC does not have a latency requirement for cellular networks,  we use $\tau$=150~ms for the UDP RTT measurements, as 100-200~ms is typically acceptable for most applications~\cite{nw_rtt}.

Our Speedtest measurements were conducted approximately every 10~minutes. We apply median censoring to determine state transitions; that is, when consecutive measurements indicate different usability states, we assume the network transitioned at the midpoint between them. 
The measurements indicate extreme performance variability, providing a rich environment for application of  QoC KPIs.
In the Appendix, Figure~\ref{fig:rw_dists}  summarizes the measured performance at each location, while Figures~\ref{speed_sqoc} and~\ref{rtt_sqoc} show the distribution of QoC KPIs for download/upload speed and RTT, respectively.


\begin{figure}[t]
  \vspace{-\intextsep}
  \centering
  \begin{subfigure}[b]{0.49\linewidth}
    \includegraphics[width=\linewidth]{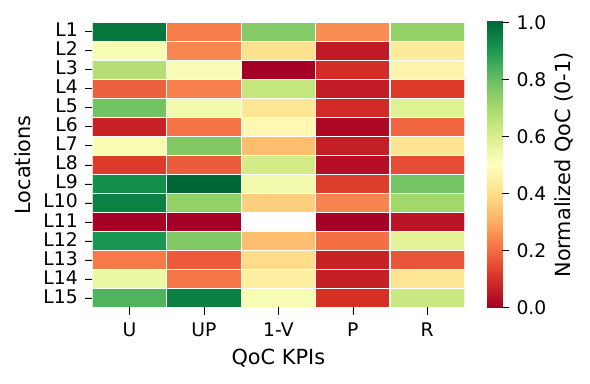}
    \caption{Throughput ($\tau$=35/3Mbps)}
    \label{rw_353}
  \end{subfigure}
  \begin{subfigure}[b]{0.49\linewidth}
is respoindin    \includegraphics[width=\linewidth]{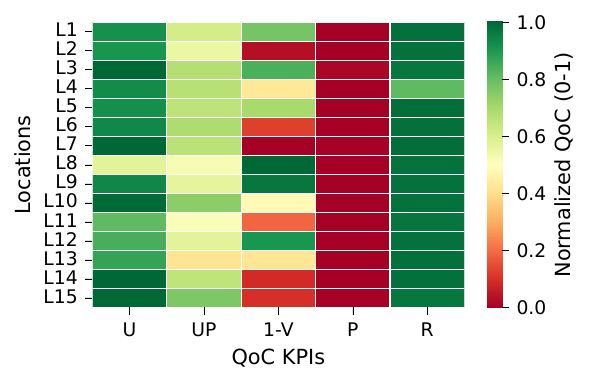}
    \caption{Latency ($\tau$=150ms)}
    \label{rw_150}
  \end{subfigure}
  \caption{QoC KPIs from AT\&T's production network.}
    \label{fig:rw_heatmaps}
\end{figure}

\begin{figure}[t]
  \vspace{-\intextsep}
  \centering
  \begin{subfigure}[b]{0.49\linewidth}
    \centering
    \includegraphics[width=\linewidth]{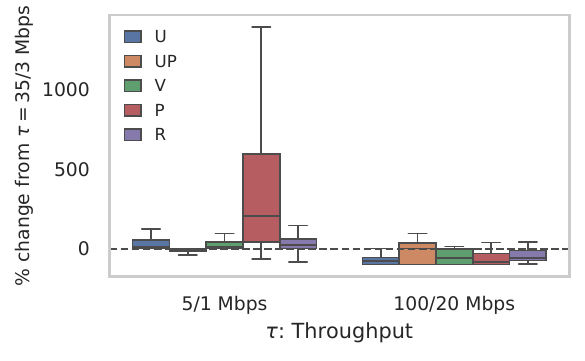}
    \caption{Throughput}
    \label{thres_st}
  \end{subfigure}
  \begin{subfigure}[b]{0.49\linewidth}
    \centering
    \includegraphics[width=\linewidth]{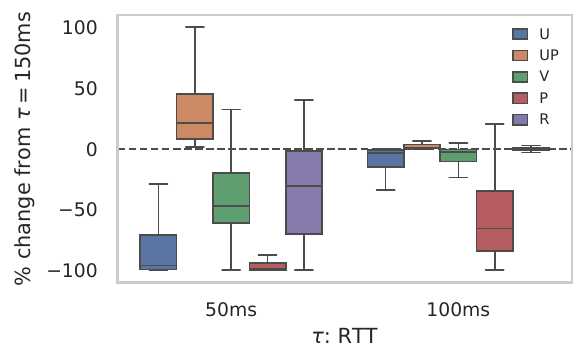}
    \caption{RTT}
    \label{thres_rtt}
  \end{subfigure}
  \caption{Responsiveness of QoC KPIs to varying $\tau$.}
    \label{fig:thresholds_qoc}
      \vspace*{-0.1in}
\end{figure}

\vspace*{-0.1in}
\subsection{What does QoC reveal about per-location network behavior?}

We present the daily average of each normalized QoC KPI as heatmaps in Figure~\ref{fig:rw_heatmaps}. 
The Speedtest results in Figure~\ref{rw_353} show substantial differences across all QoC dimensions. Several locations (L4, L6, L13) exhibit Usability below 0.25, 
while locations L1, L9, L10, and L12 achieve Usability above 0.90.  However,   the Persistence scores of these latter locations are remarkably low,  from 0.09 to 0.24. This translates to average usable durations of barely two hours before performance drops below the threshold. Resilience at these locations is  higher, indicating that while usable periods are short, the network recovers relatively quickly. Location L1 illustrates another important nuance: despite high Usability (0.96), its Usable Performance score is among the lowest (0.22), indicating that download/upload speeds consistently exceed their respective $\tau$ values, but the absolute speeds remain close to $\tau$. This demonstrates that Usable Performance and Persistence do not necessarily coincide, and understanding both dimensions is essential for holistic  coverage quality characterization.


In examination of RTT, we note that this metric is often jittery.  Hence  we apply a hysteresis~\cite{hysteresis_ip, hysteresis_cellular, hysteresis_cs} band of 10\% to reduce spurious state transitions caused by minor fluctuations around $\tau$. With this smoothing, the QoC results reveal distinct patterns,  shown in Figure~\ref{rw_150}. Most locations achieve high Usability (above 0.87), indicating that RTT is typically below 150~ms. However, Persistence is near zero and Resilience is close to one in all locations; despite being predominantly usable, RTT fluctuates wildly between usable and unusable states. Further, locations exhibit a wide range of Variability (1-V). 
For instance, L9 and L11 have comparable Usability (0.93 and 0.87), yet their 1-V scores are 1.00 and 0.00 respectively, revealing that L9 has low,  stable RTT while L11 exhibits wide RTT variation despite remaining predominantly usable.  The high Variability score at L11 indicates a likely  negative impact on latency-sensitive applications.

\vspace*{-0.1in}
\subsection{How responsive is QoC to threshold and parameter choices?}
Next, 
we  study the responsiveness of the KPI characterizations  to the specific threshold value. 
Letting $\tau$ represent download/upload speed, we evaluate additional Usability Thresholds of 5/1 and 100/20~Mbps  to understand stability properties under both more lenient and stricter application requirements. For RTT, we apply thresholds of 50~ms and 100~ms to capture responsiveness for latency-sensitive applications such as real-time gaming or video conferencing.

\noindent 
{\bf $\tau$ responsiveness: Throughput.} We compute the percentage change in each KPI relative to the baseline of 35/3~Mbps. Figure~\ref{thres_st} shows the distribution of these changes across locations. 
At $\tau$=5/1~Mbps, Persistence exhibits the largest increase, with a median change of roughly 500\% and substantial variance across locations. The remaining KPIs show modest positive changes. At $\tau$=100/20~Mbps, all KPIs decrease as expected, with Usability and Usable Performance showing the largest drops. 
The wide range of percentage change per location  
reflects  underlying differences in coverage quality. 

\noindent {\bf $\tau$ responsiveness: Latency.} 
Figure~\ref{thres_rtt} shows the percentage change relative to the 150~ms baseline for 50~ms and 100~ms thresholds. Compared to throughput, the changes across all KPIs are more moderate. At stricter  $\tau$ values, Persistence exhibits the largest decrease, as expected, followed by Variability, indicating the range of usable RTT values narrows considerably. At 
$\tau$=100~ms, all metrics other than Persistence 
remain relatively stable. These results confirm that RTT is jittery across all locations, and that the QoC KPIs respond proportionally to threshold strictness; there are modest changes at 100~ms and larger shifts at 50~ms.
Finally, we test the responsiveness of the QoC KPIs to the RTT hysteresis band. The primary impact is on Persistence; Persistence increases with wider bands as minor fluctuations around the threshold are ignored and fewer state transitions result.  
We include these results in Figure~\ref{fig:rtt_hys} in the Appendix.

\vspace*{-0.1in}
\subsection{How does QoC change local coverage conclusions?}
\label{sec:six_two}
\begin{figure}[t]
\vspace{-\intextsep}
\includegraphics[width=0.8\linewidth]{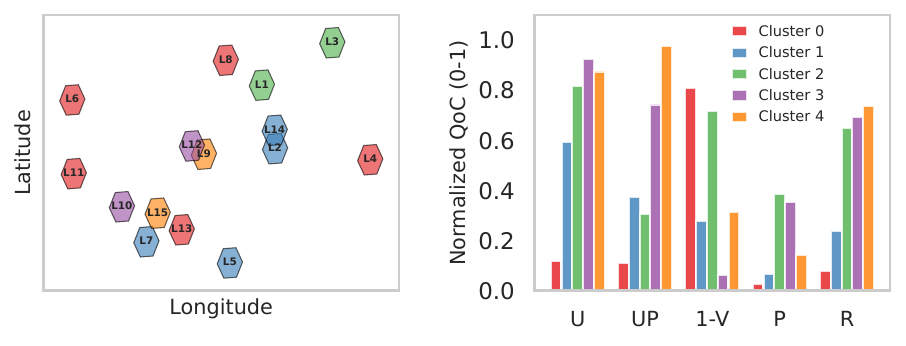}
\caption{Identification of coverage similarity. }
\label{fig:ward_rw}
\vspace*{-0.1in}
\end{figure}
We now revisit a question we posed in Section~\ref{sec:spatial}:  How homogeneous is the coverage quality within a region? To answer this question locally, we seek to understand how coverage quality varies and to identify locations with similar QoC profiles. Each location is represented by a five-dimensional vector corresponding to the normalized QoC KPIs ($U$, $M_{usable}$, $V_{usable}$, $P_{usable}$, $R$). To identify groups of locations with similar coverage characteristics, we apply Ward's hierarchical agglomerative clustering~\cite{Ward01031963}, which iteratively merges clusters by minimizing the total within-cluster variance. We select the optimal number of clusters using the silhouette score~\cite{ROUSSEEUW198753}, evaluating $K \in \{2, 3, 4, 5\}$ and selecting the value that maximizes cluster cohesion and minimizes within-cluster variability. The clustering results and corresponding QoC profiles are presented in Figure~\ref{fig:ward_rw}.

The figure shows five distinct performance clusters. Cluster~4  achieves the highest QoC across most dimensions; Cluster~0  exhibits degraded coverage, with low Usability and near-zero Persistence. This poor performance is notable because all locations are reported as 35/3~Mbps on coverage maps. Notably, the clusters are spatially interleaved rather than contiguous; neighboring locations often fall into different clusters despite geographic proximity. This spatial heterogeneity exposes the inability of traditional coverage maps to capture fine-grained variability in coverage quality. QoC identifies these localized disparities, providing actionable insights for network planners and policymakers to address coverage gaps within areas officially reported as covered.

\vspace*{-0.1in}
\subsection{How does QoC compare to current industry standards?}
To contextualize  QoC  within existing network performance characterizations, we examine how the QoC KPIs relate to other commonly used cellular network performance metrics, specifically: 
average, 5$^{th}$ and 95$^{th}$ percentile throughput, 
and Broadband Consistent Quality (BCQ).\footnote{BCQ is a metric  by OpenSignal~\cite{Opensignal} that measures the proportion of measurements that meet minimum thresholds for 
minimum thresholds for download throughput (>5Mbps), upload throughput (>1.5Mbps), latency (<50 ms), jitter (<12ms), packet loss (<1\%) and time to first byte (<0.8s). We exclude time to first byte as it is unavailable in Ookla Speedtest measurements.}
We compute the Pearson correlation between each QoC KPI and these metrics in the 15 AT\&T measurement locations; the results are presented in Figure~\ref{fig:benchmark_corr} in the Appendix. They reveal that Usability and Usable Performance exhibit strong correlations, while Persistence and Resilience show only low to moderate correlations, suggesting that temporal stability patterns remain largely absent from current metrics.

\begin{figure}[t!]
  \vspace{-\intextsep}
 \begin{subfigure}[b]{0.48\linewidth}
\includegraphics[width=\textwidth]{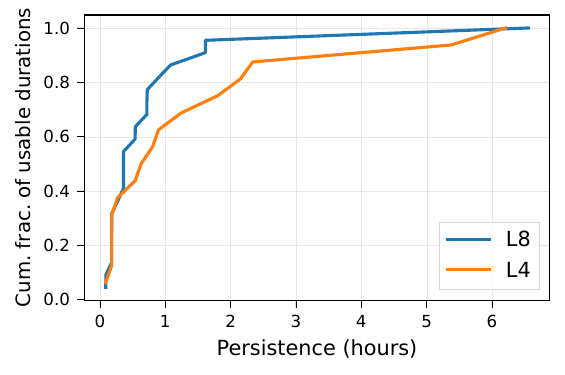}
\caption{Persistence} 
\label{fig:thresholds}
\end{subfigure}%
\begin{subfigure}[b]{0.48\linewidth}
\includegraphics[width=\textwidth]{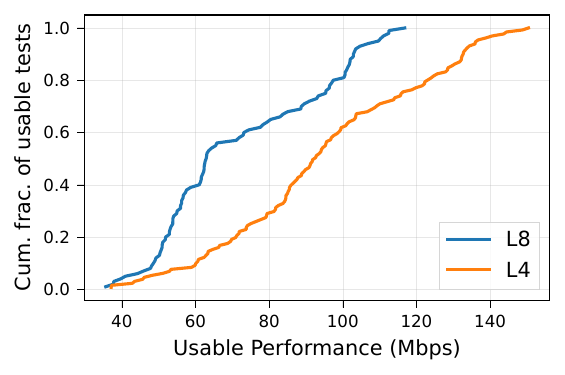}
\caption{Usable Performance} 
\end{subfigure}
\caption{Distribution of QoC KPIs for location pairs with similar average throughput.}
\label{fig:avg_bench}
\end{figure}

We next take a deeper dive to demonstrate the QoC KPIs' ability of to reveal meaningful differences in coverage quality between locations that appear equivalent with traditional metrics.
We consider location-pairs L4 and L8, which  have average download speeds of $\sim$40~Mbps and 95$^{th}$ percentile speeds 91-110~Mbps. Figure~\ref{fig:avg_bench} reveals distinct QoC profiles. L8 exhibits shorter usable durations (Persistence). 
The Usable Performance distributions diverge substantially: the median at L8 is around 60~Mbps, while that of L4 is around 100~Mbps. These differences, invisible under average throughput, could have practical implications for user experience and application performance, and QoC captures them. We find similar results for location-pairs with comparable BCQ scores, and present them in Figure~\ref{fig:avg_bcq} in the Appendix.

Next, we compare the characterizations between spatial aggregations of QoC profiles in our local community and the FCC's 80/80 consistent quality metric.\footnote{We slightly modify the 80/80 metric to report the minimum speed experienced by 80\% of locations (instead of subscribers) during 80\% of the time, during peak hours.}  To do so, we first confirm the ability of QoC spatial aggregation to retain stability differences from per-location QoC profiles. 
This is included as Figure~\ref{fig:sqoc_hex7} in the Appendix. The figure shows the spatial aggregation of QoC KPIs  at $\tau=35/3$~Mbps, revealing substantial variability across locations along all QoC dimensions.  However, 
the 80/80 metric score for the same region is 0.5~Mbps download and 0.5~Mbps upload speeds. While this single value captures the lower tail of performance, it is dominated by the poorest-performing locations in the region and misses the holistic characterization  of stability and variability. QoC captures these dimensions. 
\section{QoC Sensitivity to Data Density }
QoC benefits from dense measurements for accurate computation, which can be obtained, for example, through strategic placement of stationary measurement units by local broadband organizations. In practice, however, collecting temporally and spatially dense data is often difficult or resource-intensive. Assessing QoC’s sensitivity to data sparsity is therefore critical. Sparse data may  be available, for instance, through crowdsourced speed tests~\cite{ookla_cellmap,xlab_news,nperf_map,opensignal_map}. We evaluate this sensitivity by progressively removing data and measuring the  changes in network characterization.

\ignore{
QoC   benefits from dense measurement data for accurate computation.
Such data could be collected, e.g., by  local broadband organizations with the strategic placement of stationary measurement units at key points of interest.  More generally, however, 
temporally and spatially dense    measurements  can be difficult and/or resource-intensive to obtain. 
Thus, assessing the sensitivity of the QoC framework to  data sparsity is critical. Sparse data can be available, for instance, through crowdsourced speed tests~\cite{ookla_cellmap, xlab_news, nperf_map, opensignal_map}. 
Here, we study the sensitivity of QoC by removing different quantities of  data and evaluating the change in network  characterization. 
}

\noindent
{\bf Temporal sensitivity.}
We evaluate how reducing temporal measurement density impacts QoC KPIs at individual spatial units. 
We use two approaches to down-sample  the production network dataset: (i) fixed-interval 
and (ii) random.
For fixed, we  select one datapoint  uniformly at random at  time intervals: 10~sec, 1~min, 15~min, 1~hr, 6~hrs, and 12~hrs.
To randomly down-sample, we
irregularly sample measurements by retaining a fraction $d \in \{0.5, 0.25,0.1, 0.05, 0.01\}$ of  measurements uniformly at random without replacement.
We repeat each down-sampling experiment 30 times. 
We analyze QoC KPI fidelity loss by computing the normalized absolute errors between each down-sampled QoC KPI and its corresponding  full-data baseline.

\begin{figure}[t]
  \vspace{-\intextsep}
  \centering
  \begin{subfigure}[b]{0.48\linewidth}
    \includegraphics[width=\linewidth]{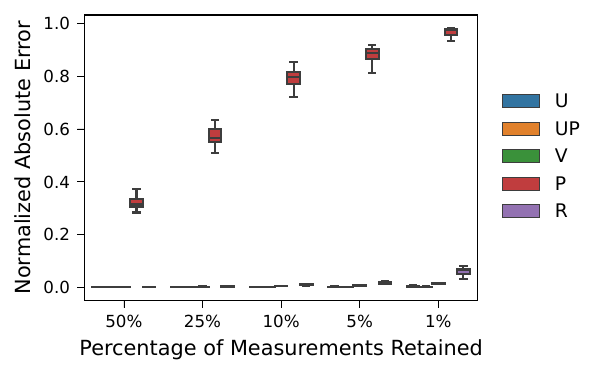}
    \caption{RTT: $\tau$=150~ms}
\label{use_periodic}
  \end{subfigure}
  \begin{subfigure}[b]{0.48\linewidth}
    \includegraphics[width=\linewidth]{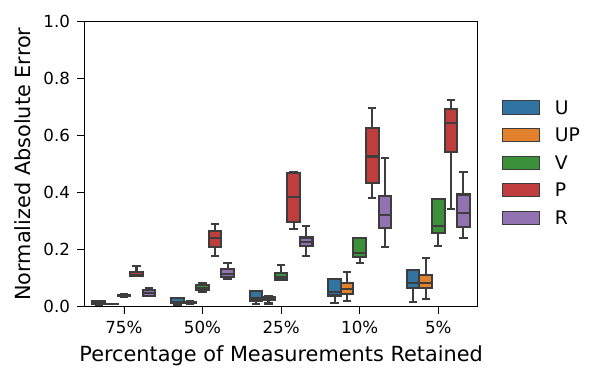}
    \caption{Throughput: $\tau$=35/3Mbps}
    \label{res_periodic}
  \end{subfigure}
  \caption{Effect of random temporal down-sampling on QoC KPIs for throughput and RTT data.}
    \label{fig:sensitivity_random}
\end{figure}


\begin{figure}[t]
  \vspace{-\intextsep}
  \centering
  \begin{subfigure}[b]{0.49\linewidth}
    \centering
    \includegraphics[width=\linewidth]{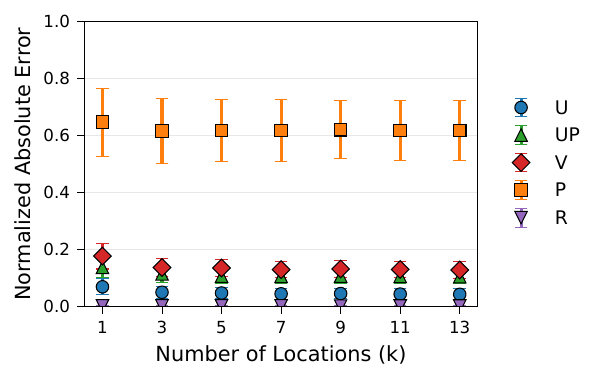}
    \caption{RTT $\tau$=150~ms, 10\%}
\label{spatial_rtt}
  \end{subfigure}
  \begin{subfigure}[b]{0.49\linewidth}
    \centering
    \includegraphics[width=\linewidth]{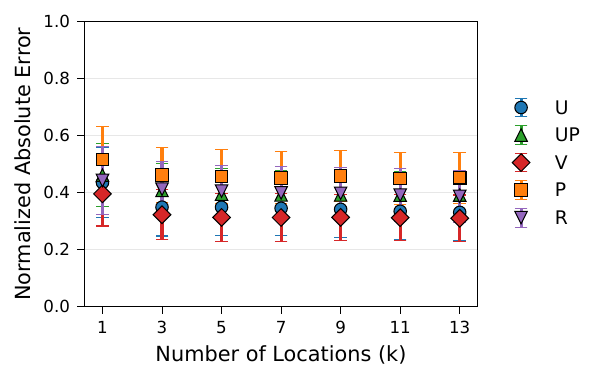}
    \caption{Throughput $\tau$=35/3~Mbps}
    \label{spatial_st}
  \end{subfigure}
  \caption{Effect of spatial down-sampling on QoC.} 
    \label{fig:sensitivity_spatial}
\end{figure}

Random down-sampling results are shown for RTT and throughput in Figure~\ref{fig:sensitivity_random}; results  are aggregated across the 15 locations. We include the results for fixed down-sampling in Figure~\ref{fig:sensitivity_fixed} in the Appendix. In both down-sampling approaches,  Persistence is consistently the most sensitive and affects RTT more than Speedtest throughput.
This sensitivity  is due to the inherent jittery behavior of RTT. Rapid fluctuations cause frequent state transitions; down-sampling therefore misses  transitions, 
leading to larger errors in Persistence estimation. 
On the other hand, Usability and Usable Performance are fairly stable in all down-sampling. Variability and Resilience show moderate sensitivity, with errors increasing gradually as measurement density decreases.

The differing sensitivity of the KPIs provides practical guidance on measurement frequency: relatively stable network behavior can be measured less often, while erratic or highly variable performance requires more frequent measurement to ensure accurate characterization.

\noindent
\textbf{Spatial sensitivity.}
We conclude our evaluation with a study of   QoC's sensitivity to spatial data sparsity. 
We  down-sample the production network dataset by reducing the number of locations aggregated into the spatial QoC estimates, 
retaining $k \in \{ 13,11,9,7,5, 3, 1 \}$ locations  of the original 15. For each $k$, we repeat the down-sampling experiment 30 times. 
We  recompute region-level QoC KPIs 
and quantify fidelity loss by calculating the normalized absolute errors between down-sampled aggregates and their corresponding full-data baselines.
The results are shown in Figure~\ref{fig:sensitivity_spatial}. For RTT, 
most KPIs show 
low error (below 0.1) even with as few as 3 locations. 
Persistence again is the outlier with higher error (0.45–0.52). 
This reflects the inherently localized nature of Persistence: temporal stability patterns vary substantially by location and cannot be entirely recovered through spatial aggregation alone. 
For Speedtests, all QoC KPIs exhibit moderate error (0.30–0.50). 
This  reflects the greater spatial heterogeneity of Speedtest QoC profiles,  which causes 
reduced data density to amplify errors when key stability patterns from constituent sub-regions are  missed. 
\vspace*{-0.1in}
\section{Related Work}
Mobile broadband assessments have primarily relied on KPIs like throughput, latency, and signal strength. 
The limitations of this narrow focus are 
emphasized in~\cite{sonntag2013mobile,gember2012obtaining}. Prior research~\cite{shakir2023key,elmokashfi2017addingthenextnine} underscores challenges such as temporal variability and sampling biases in traditional measurement approaches. Other studies~\cite{bauer2018mobilebroadband, frias2023building} 
highlight difficulties in measuring mobile broadband performance 
and emphasize the need for a 
flexible measurement framework to support evidence-based policymaking.
Recognizing limitations of throughput and latency, QoE research has developed composite models: VoIP MOS frameworks jointly measure loss, delay, and recovery~\cite{mos_1, mos_2, mos_3}; adaptive‐video QoE 
combines startup delay, bitrate variability, and rebuffering~\cite{video_qoe_1, video_qoe_2}.
Standards telecommunications bodies have set QoS benchmarks~\cite{3gpp_5g, itu_y1541}. 
Studies of QoS and QoE in 4G LTE and 5G networks  underscore the need for  metrics such as session quality and recovery~\cite{Adarsh:ICCCN21, beyond_throughput, qos_qoe_5g, qos_qoe_cellular, qos_qoe_cellular_1}.

The inherent instability and variability of network traffic have been well-documented. Paxson~\cite{Paxson1997:End} introduced  persistence and prevalence in network routing behavior.
Subsequent studies~\cite{willinger2004pragmatic,Baltrunas2014:Reliability,lazarou2009describing, reliability_imp, reliability_nationwide, bauer2012reliability, bischof2017characterizingimprovingreliabilitybroadband} documented the importance of reliability-based metrics. 
Other work~\cite{Nikravesh2014:Mobile, cellular_variability, dasari2018deviceimpact, smartphones_5g, realworld_5g, 5g_mobility, 5g_variegated, imc_23_driving, mmwave_comparison, varshika_pam, moinak_pam} has shown the variability of cellular performance
due to frequency band, device model, network load, mobility, carriers, geography and environmental conditions.

Recent studies have integrated spatio-temporal modeling into performance analyses to quantify variability in network performance~\cite{st_hotspots,st_bayesian}. 
Other work~\cite{st_deep,st_cellcenteredge} applies deep learning and clustering techniques to model and predict variations between cell-center and cell-edge users.

\noindent
{\bf FCC and industry coverage metrics.}
The FCC's Measuring Broadband America program reports an "80/80" consistent quality metric, which is the minimum speed that at least 80\% of subscribers experience at least 80\% of the time during peak periods~\cite{mba_8080}. Ookla, RootMetrics and Umlaut all use proprietary algorithms to produce composite scores based on QoS and/or QoE metrics stemming from  crowdsourced speed tests and/or drive testing~\cite{ookla_ss_method,rootmetrics_method, umlaut_bench}.
OpenSignal reports Broadband Consistent Quality (the proportion of tests meeting minimum thresholds for throughput, latency, jitter, packet loss and time-to-first-byte) and Reliability Experience (using connected time, responsiveness and task completion)~\cite{opensignal_method} from crowdsourced and drive-test-based network benchmarking that carriers and regulators use.
\section{Conclusion and Implications}
\label{sec:conclusion}

We have introduced a framework for characterizing cellular network coverage quality, usability, and stability. QoC can represent longitudinal coverage at a single location or be spatially aggregated to capture larger geographic areas. It goes beyond conventional availability metrics to provide fine-grained temporal and spatial coverage characterization. We present five QoC KPIs as a robust baseline and encourage the community to develop additional KPIs as networks and applications continue to evolve.

Collecting temporally and spatially dense measurements can be challenging. We identify which KPIs are most sensitive to data sparsity. Fortunately, quality and stability properties of cellular coverage do not need to be uniformly studied.   Critical locations, such as bus stops, downtown walking areas, urban gathering spaces, and smart city sites requiring ultra-reliable low-latency communication (URLLC), can be targeted with short-term measurement deployments (e.g., Raspberry Pis) to capture high-speed, stable coverage.

QoC KPIs can also enhance regulatory and policy processes. For example, in the FCC Mobile Challenge, stakeholders can target suspected low-performance areas and use QoC KPIs to demonstrate that provider coverage claims do not match observed service. Additionally, storing lightweight distributions of usable ($\mathcal{L}$) and unusable ($\mathcal{D}$) durations at the spatial unit level enables probabilistic queries relevant to policy, such as whether users can access reliable service long enough for critical applications like online learning or telehealth. This information can guide strategic allocation of infrastructure subsidies and empower users to choose the best local provider.

\newpage
\bibliographystyle{ACM-Reference-Format}
\bibliography{refs,refs2,belding,netgap}

\appendix
\section{Appendix}
We use the Appendix to share our  Data Release statement as well as to offer additional details about  our datasets and data aggregation methodology.  We also include  complete statistical descriptions and analyses for all QoC KPIs evaluated with our datasets.  Finally, we include auxiliary graphs and analyses not central to the core contribution of the paper but that may nonetheless be of interest to the reader.

\begin{figure}[t]
\includegraphics[width=.7\linewidth]{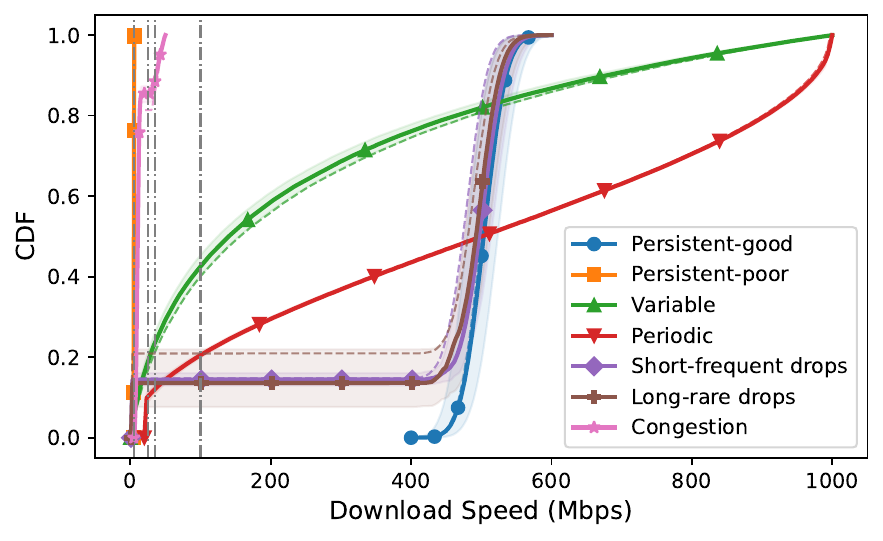}
\caption{{Distribution of download speeds in each synthetic network scenario.  Each scenario is represented by 302,400 data points.}} 
\label{fig:syn_simulations}
\end{figure}

\begin{figure*}[t!]
\centering
\begin{subfigure}[b]{0.24\linewidth}
\includegraphics[width=\textwidth]{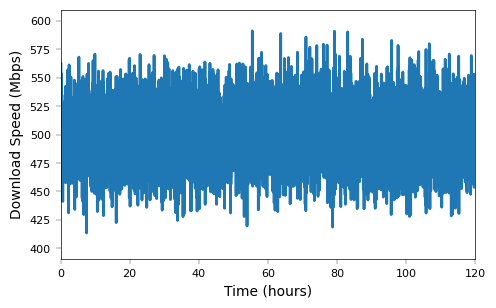}
\caption{Persistent-good\label{sim_pg}} 
\end{subfigure}%
\begin{subfigure}[b]{0.24\linewidth}
\includegraphics[width=\textwidth]{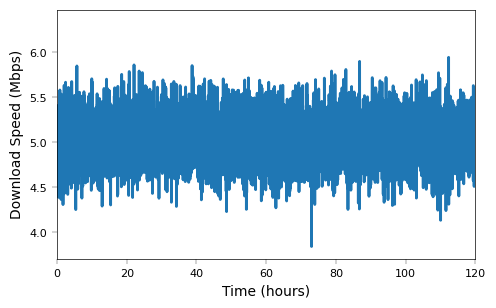}
\caption{Persistent-poor\label{sim_pp}} 
\end{subfigure}%
\begin{subfigure}[b]{0.24\linewidth}
\includegraphics[width=\textwidth]{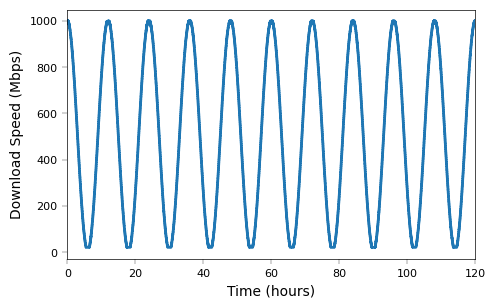}
\caption{Periodic\label{sim_per}} 
\end{subfigure}%
\begin{subfigure}[b]{0.24\linewidth}
\includegraphics[width=\textwidth]{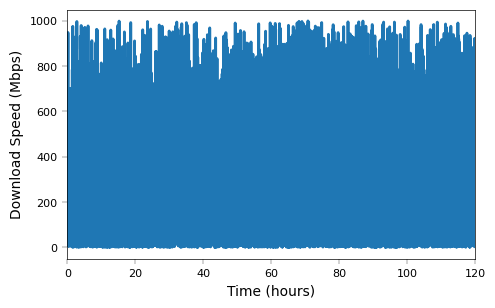}
\caption{Variable\label{sim_var}} 
\end{subfigure}
\begin{subfigure}[b]{0.28\linewidth}
\includegraphics[width=\textwidth]{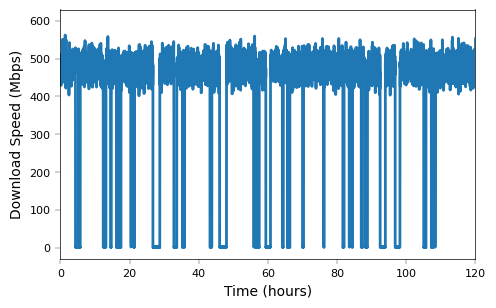}
\caption{Short-frequent drops\label{sim_sfd}} 
\end{subfigure}%
\begin{subfigure}[b]{0.28\linewidth}
\includegraphics[width=\textwidth]{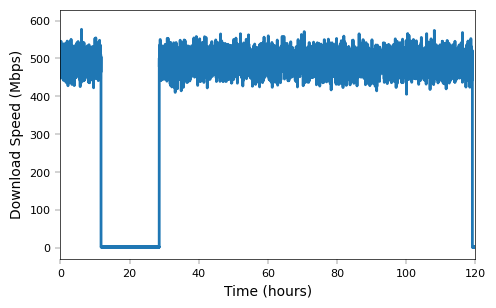}
\caption{Long-rare drops\label{sim_lrd}} 
\end{subfigure}%
\begin{subfigure}[b]{0.28\linewidth}
\includegraphics[width=\textwidth]{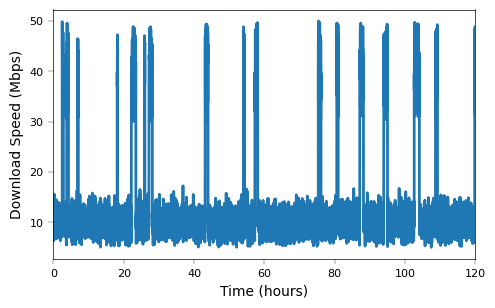}
\caption{Congestion\label{sim_con}} 
\end{subfigure}
\caption{Representative download speed time series for each synthetic scenario.} 
\label{fig:simulation_ts}
\end{figure*}

\begin{table}[t]
\centering
\small
\begin{tabular}{lp{1cm}r}
\toprule
Pairwise comparison & KS Test statistic & p-value\\
\midrule
Persistent-good vs Persistent-poor            & 1.000 & 0.000       \\
Persistent-good vs Variable                   & 0.771 & 0.000      \\
Persistent-good vs Periodic                   & 0.450 & 0.000        \\
Persistent-good vs Short-frequent drops       & 0.148 & 0.000       \\
Persistent-good vs Long-rare drops            & 0.155 & 0.000        \\
Persistent-good vs Congestion                 & 1.000 & 0.000      \\
Persistent-poor vs Variable                   & 0.940 & 0.000       \\
Persistent-poor vs Periodic                   & 1.000 & 0.000       \\
Persistent-poor vs Short-frequent drops       & 0.854 & 0.000        \\
Persistent-poor vs Long-rare drops            & 0.845 & 0.000       \\
Persistent-poor vs Congestion                 & 0.986 & 0.000      \\
Variable vs Periodic                          & 0.322 & 0.000        \\
Variable vs Short-frequent drops              & 0.625 & 0.000        \\
Variable vs Long-rare drops                   & 0.618 & 0.000       \\
Variable vs Congestion                        & 0.720 & 0.000        \\
Periodic vs Short-frequent drops              & 0.450 & 0.000       \\
Periodic vs Long-rare drops                   & 0.454 & 0.000       \\
Periodic vs Congestion                        & 0.857 & 0.000      \\
Short-frequent drops vs Long-rare drops       & 0.024 & 1.49e-73     \\
Short-frequent drops vs Congestion            & 0.854 & 0.000        \\
Long-rare drops vs Congestion                 & 0.845 & 0.000       \\
\bottomrule
\end{tabular}
\caption{KS-statistic for  synthetic network scenarios.}
\label{tab:KS_gendata}
\end{table}

\begin{table}[t]
\centering
\footnotesize
\setlength{\tabcolsep}{2pt}
\begin{tabular}{@{}p{1cm}ccccc@{}}
\toprule
\textbf{QoC Type} & \begin{tabular}[c]{@{}c@{}}No. of\\spatial units\end{tabular} & \begin{tabular}[c]{@{}c@{}}No. of\\scenarios\end{tabular} & \begin{tabular}[c]{@{}c@{}}No. of\\simulations\end{tabular} & \begin{tabular}[c]{@{}c@{}}Measurements\\per time series\end{tabular} & \begin{tabular}[c]{@{}c@{}}Total\\data points\end{tabular} \\
\midrule
Temporal & 1 & 7 & 50 & 43,200 & 15,120,000 \\
Spatial  & 7 & 7 & 50 & 43,200 & 105,840,000 \\
\bottomrule
\end{tabular}
\caption{Details of  synthetic dataset.}
\label{tab:datagen_breakdown}
\vspace*{-0.25in}
\end{table}

\subsection{Synthetic Dataset Details}

Figure~\ref{fig:syn_simulations} shows the download speeds for each of the synthetic network scenarios. Here we provide additional details about these scenarios.

\noindent
\textbf{Persistent-good (PG):}
 This scenario models a network that demonstrates consistent download speeds 
 (i.e., temporal performance stability) well above the broadband speed threshold. For each hex10 hexagon, download speeds are drawn independently from a normal distribution of mean 500~Mbps with a small random offset ($\pm$5\%)  and hard bounds between 400 and 600~Mbps, 
 with a coefficient of variation $(\mathrm{CV})=0.05$. 

\noindent
\textbf{Persistent-poor (PP):} 
Using the same  $\mathrm{CV}=0.05$  but shifting the mean download speed to 5~Mbps  with a small random offset ($\pm$5\%) and hard bounds between 1 and 20~Mbps,
this scenario models a temporally stable low performance network. 


\noindent
\textbf{Periodic:}  
This scenario models networks with diurnally varying performance, such as an environment influenced by cyclical demand (e.g., cell load). For each hex10 hexagon, we construct a sinusoidally modulated mean centered around a baseline download speed of 500Mbps, and an amplitude of 500Mbps with a small random offset ($\pm$1\%) applied to both the base mean and the amplitude;  hard bounds between 1 and 1000~Mbps. The resulting time-varying mean is given by $\mu(t) = \mu_{\mathrm{base}} + A \cdot \cos(4\pi t / 1440)$;  $t$ denotes the minute-level timestamp over a 24-hour period. 
This produces  periodic fluctuations in performance while preserving controllable variance across time.


\noindent
\textbf{Variable:} 
Empirical studies indicate that typical network traffic measurements are right skewed and closely follow a log-normal distribution~\cite{Alasmar2021:LogNormal}. Hence, we use log-mean values of $500$ Mbps for download speeds with a small random offset ($\pm$5\%), and a high sigma $\mathrm{\sigma}=2.5$ to get right-skewed behavior. This scenario  evaluates a network characterized by more realistic performance variability.

\noindent
\textbf{Reduced network usability:}
In some circumstances, users may experience time periods in which the network is "down" and hence inaccessible, or where it is so poorly performing that it is essentially unusable.
As described in Section~\ref{sec:synthetic}, we utilize a two-state Hidden Markov Model to simulate extended periods of unusable performance.  The details are as described in that section. 
\ignore{
, we use a two-state Hidden Markov Model (HMM). The hidden variable has two values: state 0 (usable) and state 1 (unusable). At each time step, the HMM selects  the network  state.
The state transition matrix is defined as:
\begingroup
\vspace{-0.02in}
\setlength\arraycolsep{2pt}
  \small                  
  \[
 A = \bigl(\begin{smallmatrix}
1 - p_{\mathrm{entry}} & p_{\mathrm{entry}}\\[-2pt]
1 - p_{\mathrm{self}} & p_{\mathrm{self}}
\end{smallmatrix}\bigr),
\; p_{\mathrm{entry}} = P(s_t{=}1\mid s_{t-1}{=}0),
\; p_{\mathrm{self}} = P(s_t{=}1\mid s_{t-1}{=}1).
\]
\endgroup

\noindent
This model controls how often performance drops begin ($p_{\mathrm{entry}}$) and how long they persist ($p_{\mathrm{self}}$). Within each state, we simulate download speed values using a normal distribution. The usable and unusable states are described in Table~\ref{tab:qoc-scenarios}. 
We simulate the following network types.  In the first two, 
 the network is expected to be unusable for the same total duration (15\% of total time), but the frequency and duration of disruptions vary.
}

\textbullet\ \textbf{Short-frequent drops (SFD):}  
This scenario models short, frequent disruptions, with $p_{\mathrm{entry}} = 1/180$ and $p_{\mathrm{self}} = 1 - 1/30$. This results in average degraded durations of $\sim$30 minutes and usable periods of $\sim$3 hours. 

\textbullet\ \textbf{Long-rare drops (LRD):}  
This scenario models infrequent but longer-lasting disruptions with $p_{\mathrm{entry}} = 1/4320$ and $p_{\mathrm{self}} = 1 - 1/720$. This results in degraded durations of $\sim$12 hours, occurring once every 3 days on average. 

\textbullet\ \textbf{Congestion:}
This scenario models chronic slowdowns punctuated by occasional relief using normal distributions;  
$p_{\mathrm{entry}} = 1/360$ and $p_{\mathrm{self}} = 1 - 1/60$. This results in average degraded durations of $\sim$6 hours and usable periods of $\sim$60 minutes.

The temporal aspects of these synthetic datasets are illustrated with time series representations of download speeds  in Figure~\ref{fig:simulation_ts}.
In Table~\ref{tab:KS_gendata}, we report the two-sample Kolmogorov-Smirnov (KS) test statistic comparing the distributions of download speeds generated for each  scenario to demonstrate their statistical difference. The results confirm that the distributions of each network scenario statistically  diverge from each other. 
In Table~\ref{tab:datagen_breakdown}, we summarize details about the number of time series measurements we use for each QoC KPI characterization analysis.

\subsection{Supplemental Data and Results}

\begin{table}[htbp]
  \centering
  \caption{Behavior of QoC KPIs as the Usability Threshold $\tau$ varies.}
  \label{tab:qoc_properties}
  \small
  \begin{tabular}{@{}p{0.6cm}p{6.4cm}@{}}
    \toprule
    \textbf{KPI} & \textbf{Behavior as $\tau$ varies} \\ 
    \midrule
    $U$ & Monotonically non-increasing with stricter thresholds.\\
    $UP$ & Monotonically non-decreasing with stricter $\tau$.\\
    $P$ & $P_{max}$ is monotonically non-decreasing with lenient $\tau$, but averages may fluctuate due to fragmented usable periods.\\
    $V$ & Not monotonic.\\
    $R$ & $R_{min}$ (or maximum recovery time $D_{max}$) is monotonically non-increasing with lenient $\tau$, but averages may fluctuate.\\
    \bottomrule
  \end{tabular}
\end{table}

\begin{figure}[t]
\centering
 \begin{subfigure}[b]{0.49\linewidth}
\includegraphics[width=\textwidth]{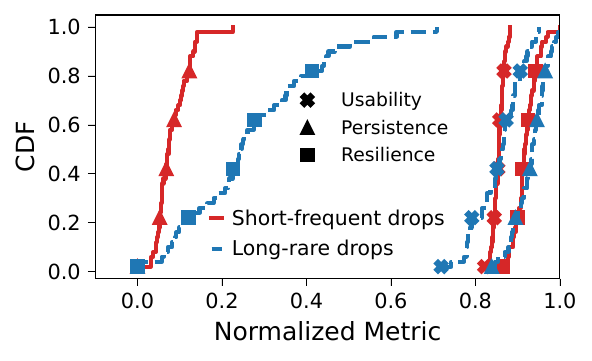}
    \caption{Different drop types}
    \label{fig:drops}
    \end{subfigure}
     \begin{subfigure}[b]{0.49\linewidth}
\includegraphics[width=\textwidth]{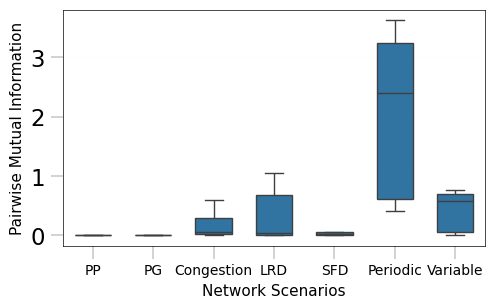}
  \caption{Mutual information}
  \label{fig:mut_info}
    \end{subfigure}
    \caption{QoC KPIs under different drop types and their mutual information for synthetic data.}
\end{figure}


\begin{figure*}[t]
  \vspace{-\intextsep}
  \centering
  \begin{subfigure}[b]{0.24\linewidth}
    \centering
    \includegraphics[width=\linewidth]{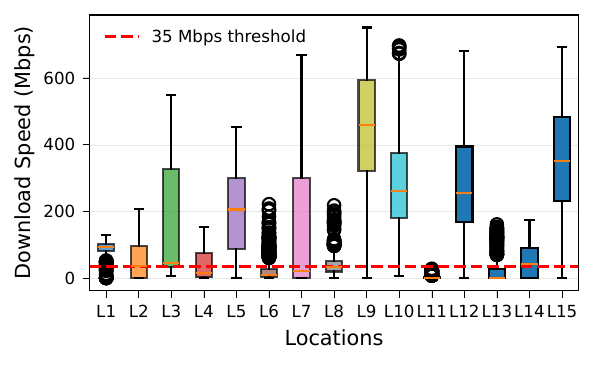}
    \caption{Download Speed}
    \label{rw_dl}
  \end{subfigure}
  \begin{subfigure}[b]{0.24\linewidth}
    \centering
    \includegraphics[width=\linewidth]{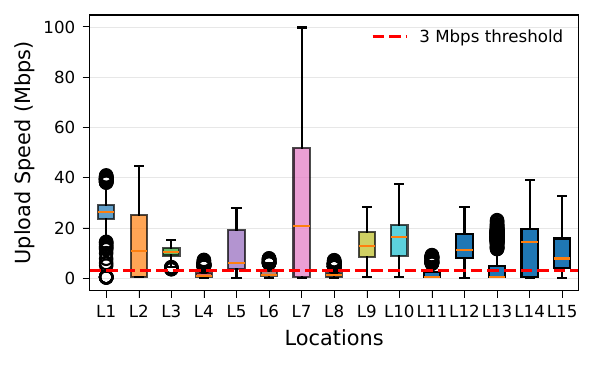}
    \caption{Upload Speed}
    \label{rw_ul}
  \end{subfigure}
  \begin{subfigure}[b]{0.24\linewidth}
    \centering
    \includegraphics[width=\linewidth]{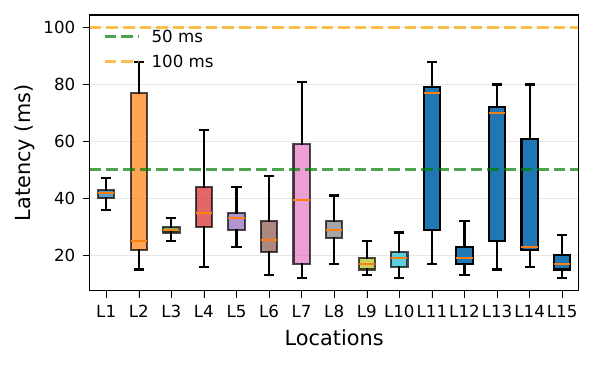}
    \caption{Ookla Latency}
    \label{rw_lat}
  \end{subfigure}
  \begin{subfigure}[b]{0.24\linewidth}
    \centering
    \includegraphics[width=\linewidth]{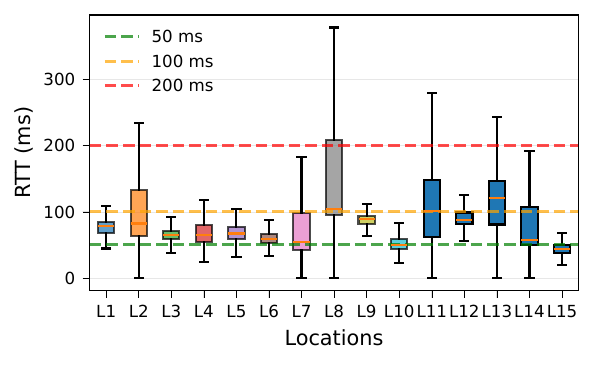}
    \caption{UDP RTT}
    \label{rw_rtt}
  \end{subfigure}
  \caption{Distribution of Speedtest measurements and experimental UDP RTT measurements across the 15 measurement locations.}
    \label{fig:rw_dists}
\end{figure*}

\begin{figure*}[t]
  \vspace{-\intextsep}
  \centering
  \begin{subfigure}[b]{0.49\linewidth}
    \centering
    \includegraphics[width=\linewidth]{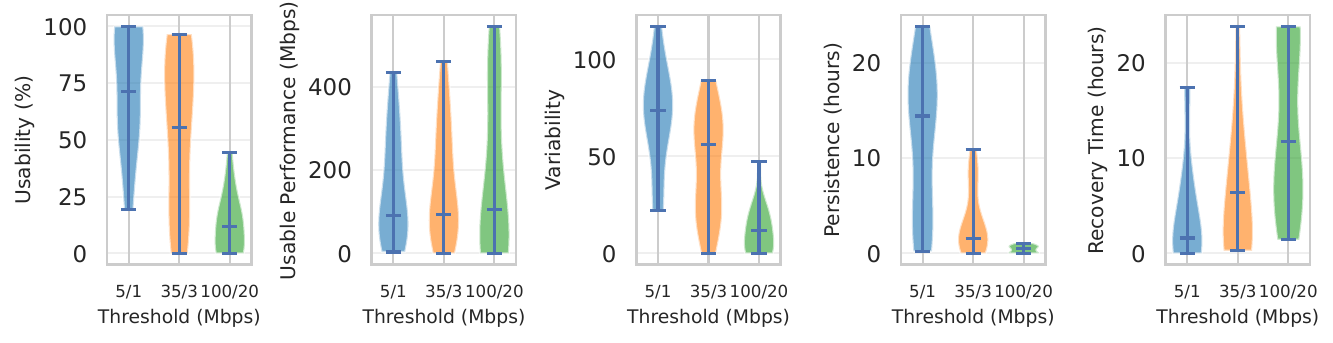}
    \caption{Throughput}
    \label{speed_sqoc}
  \end{subfigure}
  \begin{subfigure}[b]{0.49\linewidth}
    \vspace{-\intextsep}
    \centering
    \includegraphics[width=\linewidth]{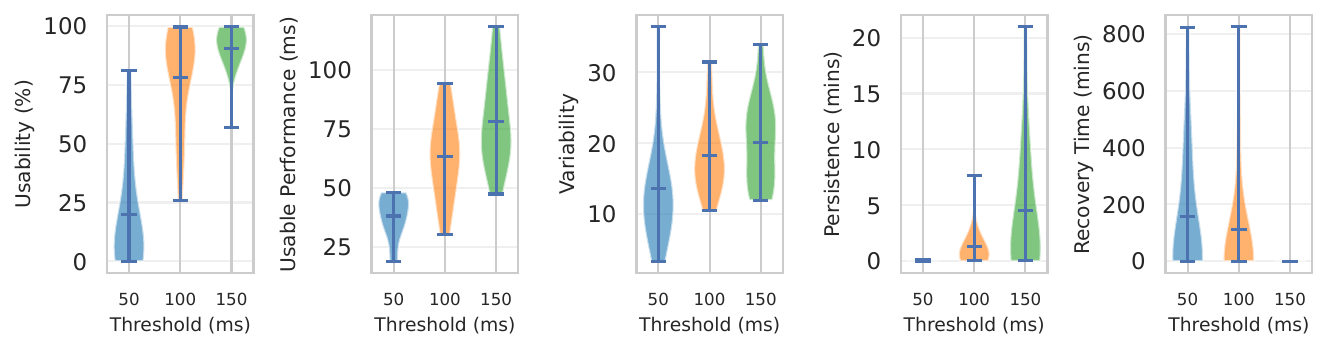}
    \caption{RTT}
    \label{rtt_sqoc}
  \end{subfigure}
  \caption{QoC distributions for real-world throughput and RTT measurements.}
    \label{fig:qoc_violin}
\end{figure*}

\begin{figure}[t]
    \centering
    \includegraphics[width=0.5\linewidth]{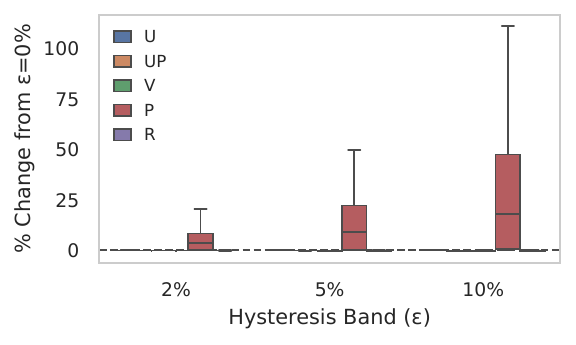}
    \caption{Effect of RTT hysteresis bands on QoC KPIs.}
\label{fig:rtt_hys}
\end{figure}



\begin{figure}[t!]
\includegraphics[width=0.9\linewidth]{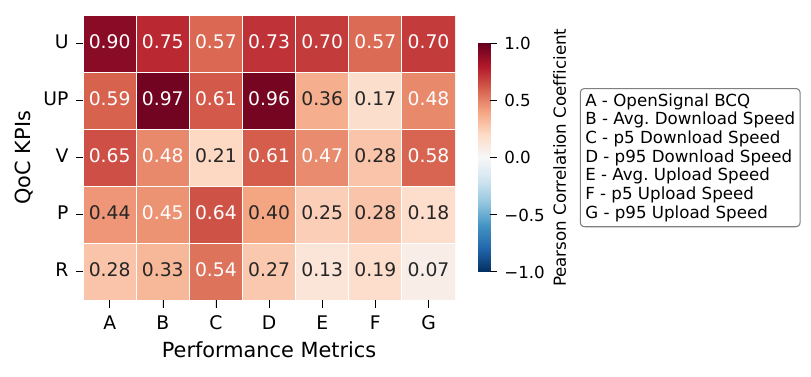}
    \caption{QoC KPIs vs. standard mobile coverage KPIs.}
    \label{fig:benchmark_corr}
\end{figure}

\begin{figure}[t]
\centering
\begin{subfigure}[b]{0.48\linewidth}
\includegraphics[width=\textwidth]{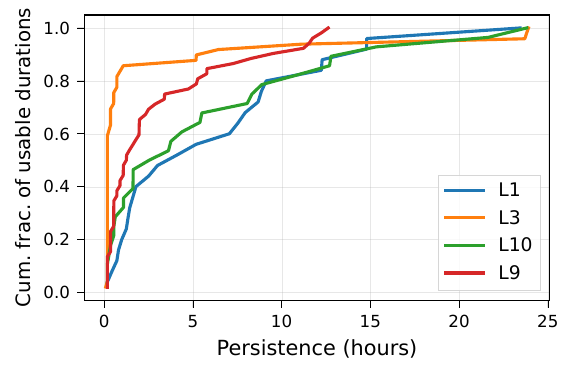}
\caption{Persistence} 
\label{fig:thresholds}
\end{subfigure}%
\begin{subfigure}[b]{0.48\linewidth}
\includegraphics[width=\textwidth]{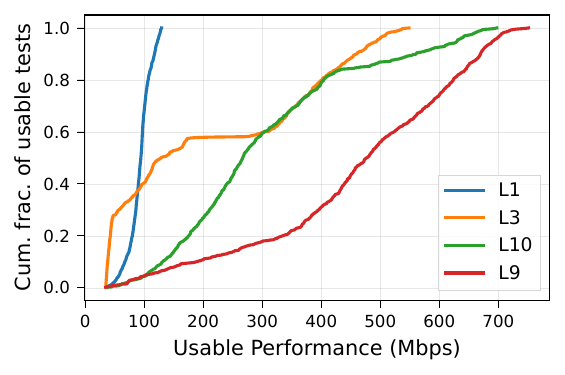}
\caption{Usable Performance} 
\end{subfigure}
\caption{Distribution of QoC KPIs for location pairs with similar Broadband Consistent Quality scores.}
\label{fig:avg_bcq}
\end{figure}

\begin{figure}[t]
  \vspace{-\intextsep}
  \centering
  \begin{subfigure}[b]{0.37\linewidth}
    \includegraphics[width=\linewidth]{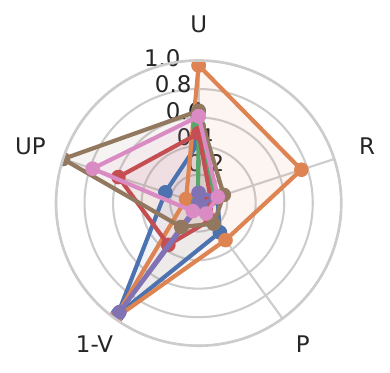}
    \caption{Throughput:$\tau=$35/3Mbps}
    \label{speed_sqoc}
  \end{subfigure}
  \begin{subfigure}[b]{0.61\linewidth}
    \includegraphics[width=\linewidth]{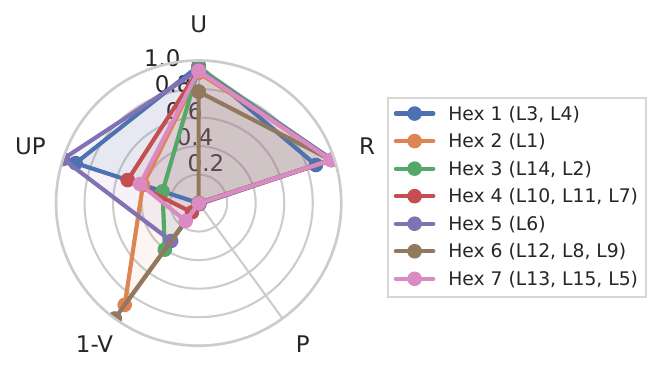}
    \caption{RTT: $\tau=$150ms;1 0\%}
    \label{rtt_sqoc}
  \end{subfigure}
  \caption{Spatial QoC for AT\&T measurements.}
    \label{fig:sqoc_hex7}
\end{figure}

\begin{figure}[t]
  \vspace{-\intextsep}
  \centering
  \begin{subfigure}[b]{0.48\linewidth}
    \includegraphics[width=\linewidth]{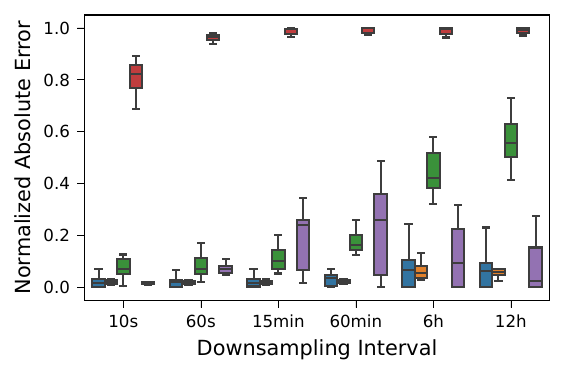}
    \caption{RTT: $\tau$=150 ms}
\label{use_periodic}
  \end{subfigure}
  \begin{subfigure}[b]{0.48\linewidth}
    \includegraphics[width=\linewidth]{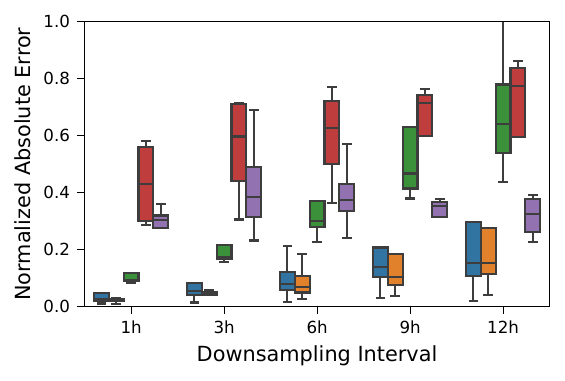}
    \caption{Throughput: $\tau$=35/3 Mbps}
    \label{res_periodic}
  \end{subfigure}
  \caption{Effect of fixed-interval temporal down-sampling on QoC KPIs.}
    \label{fig:sensitivity_fixed}
\end{figure}


\subsubsection{QoC KPI characterization and spatial evaluation.}

We present the expected behavior of each QoC KPI as the Usability Threshold~$\tau$ varies in Table~\ref{tab:qoc_properties}. 
In Figure~\ref{fig:drops}, we closely examine the distributions of Usability, Persistence and Resilience of the SFD and LRD networks.  Recall that these networks are both unusable for 15\% of the time, but the frequency and duration of disruptions vary. While average Usability
is the same,\footnote{A Wasserstein statistical distance of only 0.035.} Persistence and Resilience differ significantly; LRD networks  have  higher Persistence since  intervals of unusability are infrequent, while frequent periods of unusability in  SFD networks fragment continuity more often. Resilience is higher for SFD networks  because the network is able to quickly recover from  short unusable periods, while the long disruptions in the LRD networks require more time to recover, leading to lower Resilience. 
In Figure~\ref{fig:mut_info}, we present the distribution of pair-wise mutual information across all QoC KPIs. Mutual information is low and varies depending on the type of scenario, implying that the QoC KPIs capture different aspects of temporal continuity and are not redundant.

\subsubsection{QoC applied to production cellular networks.}
To supplement our observations in Section~\ref{sec:production}, we present the distributions of Ookla Speedtest download speeds, upload speeds, latency, as well as our experimental UDP RTT to aour university-owned server in Figure~\ref{fig:rw_dists}. Speeds and RTT values are highly variable across locations. Ookla latency and  UDP RTT are highly correlated; however, the UDP RTT values are higher, potentially because Ookla latency values are minimum latency values from ICMP pings. In Figure~\ref{fig:qoc_violin}, we present the distribution of QoC KPIs for throughput and RTT across all 15 measurement locations for three $\tau$ values. There is substantial variability across all QoC KPIs, with Usability and Persistence decreasing sharply at stricter thresholds.

Figure~\ref{fig:rtt_hys} shows the effect of different RTT hysteresis bands on QoC KPIs. Persistence increases consistently with larger hysteresis bands; medians reach approximately 50\% higher at 10\% hysteresis. This is expected: wider hysteresis bands filter out minor fluctuations around the threshold, resulting in fewer state transitions and longer sustained usable periods. The remaining KPIs are relatively stable, indicating that hysteresis primarily affects temporal persistence without substantially altering other aspects of coverage quality. 

To understand the relationship between QoC and commonly used cellular network performance metrics, we compute the Pearson correlation between each QoC KPI and average, 5$^{th}$ and 95$^{th}$ percentile throughput,
and Opensignal's Broadband Consistent Quality (BCQ) in the 15 AT\&T measurement locations; the results are presented in Figure~\ref{fig:benchmark_corr}. They reveal that Usability and Usable Performance exhibit strong correlations, while Persistence and Resilience show only low to moderate correlations, suggesting that temporal stability patterns remain largely absent from current metrics.

Finally, we demonstrate the ability of the QoC KPIs to reveal meaningful differences in coverage quality between locations that appear equivalent under similar traditional metrics.  
To demonstrate this, we identify that locations L1, L3, L9, and L10  achieve comparable Broadband Consistent Quality scores, as shown in Figure~\ref{fig:avg_bcq}, yet, their QoC profiles diverge substantially. L1 exhibits the shortest usable durations, with 80\% of usable periods lasting less than 5 hours, while L9 sustains usable periods extending to 24 hours. Recovery times also differ: L1 and L3 recover within 2 hours for nearly all unusable periods, whereas L10 experiences occasional recovery times up to 8 hours. The Usable Performance distributions reveal the starkest contrast: L1 clusters tightly between 50-150~Mbps, while L9 and L10 span 300-750~Mbps. These patterns, indistinguishable with the BCQ metric, could result in markedly different user experiences. 

\subsubsection{Spatial aggregation of QoC profiles.}
Here, we confirm the ability of QoC spatial aggregation to retain stability differences from per-location QoC profiles, using the AT\&T production network data. To do so, we spatially aggregate  the QoC profiles at each location into larger spatial representations.\footnote{Each  location belongs to a separate H3 resolution-9 hexagon. At resolution-7, the 15 locations form seven distinct hexagons, each containing one or more measurement locations. We perform our spatial aggregation on these 7 hexagons.}  
We present the findings for throughput and RTT in Figure~\ref{fig:sqoc_hex7}. The QoC profiles of each hexagon differ and reflect the combined effect of their constituent measurement locations, as was shown in Figure~\ref{rw_353}, demonstrating that the aggregation preserves localized stability patterns while enabling regional characterization.

\subsubsection{Sensitivity of QoC framework.}
We present the findings of our temporal sensitivity analysis on the QoC KPIs under fixed-interval temporal down-sampling, in Figure~\ref{fig:sensitivity_fixed}. Consistent with  random down-sampling, we find that Persistence is the most sensitive KPI. 
\end{document}